\pdfoutput=1
\documentclass[sigconf]{acmart}
\usepackage{algorithm2e}
\usepackage{multirow}
\usepackage{enumitem}

\newcounter{JSNumberOfComments}
\stepcounter{JSNumberOfComments}

\newcounter{HWNumberOfComments}
\stepcounter{HWNumberOfComments}

\AtBeginDocument{%
  \providecommand\BibTeX{{%
    \normalfont B\kern-0.5em{\scshape i\kern-0.25em b}\kern-0.8em\TeX}}}

\copyrightyear{2023} 
\acmYear{2023} 
\setcopyright{acmlicensed}\acmConference[WebSci '23]{15th ACM Web Science Conference 2023}{April 30-May 1, 2023}{Evanston, TX, USA}
\acmBooktitle{15th ACM Web Science Conference 2023 (WebSci '23), April 30-May 1, 2023, Evanston, TX, USA}
\acmPrice{15.00}
\acmDOI{10.1145/3578503.3583606}
\acmISBN{979-8-4007-0089-7/23/04}

\RestyleAlgo{ruled}
\begin{document}

%%
%% The "title" command has an optional parameter,
%% allowing the author to define a "short title" to be used in page headers.
\title{Wearing Masks Implies Refuting Trump?: Towards Target-specific User Stance Prediction across Events in COVID-19 and US Election 2020}
\renewcommand{\shorttitle}{Wearing Masks Implies Refuting Trump?}
%%
%% The "author" command and its associated commands are used to define
%% the authors and their affiliations.
%% Of note is the shared affiliation of the first two authors, and the
%% "authornote" and "authornotemark" commands
%% used to denote shared contribution to the research.

% \author{Anonymous}

\author{Hong Zhang}
%\authornote{Both authors contributed equally to this research.}
\email{tommyzhanghong@gmail.com}
\orcid{0000-0003-1193-0337}
\affiliation{%
  \institution{Singapore Management University}
  %\streetaddress{P.O. Box 1212}
  %\city{Singapore}
  \country{Singapore}
  \postcode{178903}
}

\author{Haewoon Kwak}
\email{haewoon@acm.org}
\orcid{0000-0003-1418-0834}
\affiliation{%
  \institution{Indiana University Bloomington}
  %\streetaddress{P.O. Box 1212}
  \city{Bloomington, IN}
  \country{USA}
  \postcode{47408}
}

\author{Wei Gao}
\email{weigao@smu.edu.sg}
\affiliation{%
  \institution{Singapore Management University}
  %\streetaddress{P.O. Box 1212}
  %\city{Singapore}
  \country{Singapore}
  \postcode{178903}
}

\author{Jisun An}
\email{jisun.an@acm.org}
\orcid{0000-0002-4353-8009}
\affiliation{%
  \institution{Indiana University Bloomington}
  %\streetaddress{P.O. Box 1212}
  \city{Bloomington, IN}
  \country{USA}
  \postcode{47408}
}

%%
%% By default, the full list of authors will be used in the page
%% headers. Often, this list is too long, and will overlap
%% other information printed in the page headers. This command allows
%% the author to define a more concise list
%% of authors' names for this purpose.
% \renewcommand{\shortauthors}{Trovato and Tobin, et al.}

%%
%% The abstract is a short summary of the work to be presented in the
%% article.
\begin{abstract}
    % The study of human behavior has mostly focused on understanding and explaining human behaviors online and offline. However, little is done to predict how a person would behave given his actions in related events. 
    People who share similar opinions towards controversial topics could form an echo chamber and may share similar political views toward other topics as well. The existence of such connections, which we call \emph{connected behavior}, gives researchers a unique opportunity to \textit{predict} how one would behave for a future event given their past behaviors. 
    In this work, we propose a framework to conduct connected behavior analysis. Neural stance detection models are trained on Twitter data collected on three seemingly independent topics, i.e., wearing a mask, racial equality, and Trump, to detect people's stance, which we consider as their online behavior in each topic-related event. 
    % Since a large amount of labeled training data is costly to obtain, we propose a hashtag-based stance labeling method to annotate millions of training data with very little manual effort required. We validate the quality of the stance obtained using social network graph visualization. 
    Our results reveal a strong connection between the stances toward the three topical events and demonstrate the power of past behaviors in predicting one's future behavior. 
    % the potential of connected behavior research. 
    % We also conduct an ablation study to measure the impact of  profiles, Twitter statistics, content, and stance features on prediction, and find that stances toward other events are a strong predictor. 
\end{abstract}

%%
%% The code below is generated by the tool at http://dl.acm.org/ccs.cfm.
%% Please copy and paste the code instead of the example below.
%%


%%
%% Keywords. The author(s) should pick words that accurately describe
%% the work being presented. Separate the keywords with commas.

%% A "teaser" image appears between the author and affiliation
%% information and the body of the document, and typically spans the
%% page.

% \received{20 February 2007}
% \received[revised]{12 March 2009}
% \received[accepted]{5 June 2009}

%%
%% This command processes the author and affiliation and title
%% information and builds the first part of the formatted document.
\maketitle

\section{Introduction}

% \hw{@all: Our current flow could be improved; we start from COVID, move on to the US election, move on to three topics (but no mention of racial equality before), then proceed to the connected behavior, and again explain COVID as a case study, etc.}

COVID-19 unfolded in late 2019 and rapidly spread to the whole world. In the following two years, the virus caused more than 860,000 deaths in the US and infected more than 69 million people. It became one of the most severe economic, political, and social catastrophes that impacted countless businesses and lives. As the number of cases surged in the US since its first occurrence, it quickly became a politicized health crisis~\cite{young2022politics}.
% , young2020ideological}. 

According to a survey by Pew Research~\cite{van2020facts}, 78\% of Democrats believed that COVID-19 was a major threat to the health of the US population, whereas only 52\% of Republicans shared the same view. 
Donald Trump, the 45th president of the United States from 2017 to 2021, made public announcements on multiple occasions stating that he would not wear a mask despite the US Centers for Disease Control and Prevention (CDC) advising doing so. 
Trump became an icon for anti-mask, and his action spurred the same sentiment toward wearing masks for many Republican supporters. 
Partisanship was one of the most important factors for mask adoption; the willingness to wear a mask is significantly lower in counties where Trump found strong support during the 2016 presidential election~\cite{kahane2021politicizing}. %milosh2020political,
Many studies found political influence on mask adoption as a response to COVID-19 pandemic~\cite{young2022politics, shin2022mask, adolph2022governor}.

% \tz{tommy: Added a paragraph below to fill in the gap about racial equality}

On May 25, 2020, Minneapolis police officer arrested George Floyd, a 46-year-old black man, following a 911 call reporting Mr. Floyd bought cigarettes from a convenience store with a counterfeit bill. Unfortunately, a series of violent and fatal actions by officers left Mr. Floyd unable to breathe, leading to his death. The tragic death of George Floyd caused social unrest across the United States, and tens of thousands of people swarmed the streets to express their outrage toward racial inequality~\cite{derrick2021trump}. A study had shown that Trump was unconcerned about racial inequality; he actually appeared to set out to exacerbate and inflame the issues in the United States~\cite{mcclain2021trump}. A poll in 2020 also found that two-thirds of Americans think Trump has increased racial tensions following the death of George Floyd~\cite{domenico2020trump}.
Mask adoption and racial inequalities are not the only politicized social issues. Recent studies found that opinions about gun control~\cite{mejova2022modeling} and hate towards Asians during COVID-19~\cite{an-etal-2021-predicting-anti} are also shown to be politicized. 

We increasingly observe that some opinions and behaviors are linked to each other. This could be due to the existence of ``echo chambers'' \cite{cinelli2021echo, del2016spreading,colleoni2014echo} on social media.
Social media users with a certain view on a topic tend to post and read content from people having a similar view~\cite{an2014partisan}. This creates echo chambers where users favoring content adhering to their beliefs form online communities with a shared narrative. 
People within an echo chamber expose themselves to information reinforcing their own opinions, which could facilitate social extremism~\cite{lee2022storm} and political polarization~\cite{barbera2015tweeting}. 
As such, people who share similar opinions towards controversial topics could form an echo chamber and share similar political views toward other topics as well.

The existence of such connections, which we call \emph{connected behavior}, gives researchers a unique opportunity to \textit{predict} how one would behave for a future event given their past behaviors. 
Such ability to explain and predict future behavior could bring immense value to both the scientific community and the business world. 
Despite the unprecedented body of work done on understanding online opinion and behavior based on social media data~\cite{mahmud2016predicting, bing2014public,kallus2014predicting,o2010tweets}, little was done on the ability to predict one's behavior given their behaviors in related events.

Our goal is to investigate how one's behavior in a target event is associated with his behaviors in other related events (i.e., \emph{connected behavior}) and, ultimately, predict one's future behavior given his previous behaviors. 
As a first step to studying connected behavior, this study considers opinion expression as online behavior. 
In particular, we examine individuals' stances on  three seemingly independent but related controversial topics during COVID-19: wearing masks, racial equality, and Donald Trump. 
These topics correspond to one's behavior in three events in 2020: mask adoption, racial unrest, and US Election. 
We adopt the stance detection technique to effectively and accurately identify stances towards selected topics in the study population. Stance detection has been a popular task studied by many NLP researchers in recent years; it detects a user’s opinion toward a specific target. Targets range from abstract ideas to concrete entities and events~\cite{ranade2013stance}.
            
In this study, we collect and study tweets on the three selected controversial topics and classify users' stances towards each topic using NLP techniques and neural network models. We then analyze users' stances on three topics to uncover any underlying relationship and validate the theory of connected behavior. 
Our contributions are summarized as follows:
\begin{itemize}[leftmargin=*]
    \item We empirically validate connected behavior on three selected topics, which can be helpful for researchers in the field of political and social science. We believe this is the first-of-its-kind study that examines the existence of connected behavior in a data-driven way, going beyond self-reported surveys and small-scale experiments.
    \item We propose a novel two-step stance detection framework that allows for a fast and effective way of analyzing connected behavior on multiple topics without having costly labeled data.
    \item We build a model that predicts user stance on an event based on his stances on the related events and compares it with the models based on other content-based features. We demonstrate the predictive power of stance as compared to other features. 
    \item A large dataset is collected for three controversial topics to study social polarization and NLP-related tasks. The data is available at \url{https://github.com/tommyzhanghong/connected_behavior}.
\end{itemize}

\section{Background}

The United States presidential election 2020 was an unprecedented event. The incumbent president, Republican Donald Trump, was defeated by the former 47th vice president of the United States, Democrat Joe Biden.
% \cite{wallenfeldt_2021}. 
The 2020 election had the highest number of voter turnout in the entire history of the United States, casting nearly 158.4 million ballots \cite{desilver_2021}. 
Even after the election results were announced, Trump acknowledged Biden had won but refused to concede the election. Trump claimed the election was rigged and launched a slew of lawsuits in several states that he had lost, but was unsuccessful~\cite{bbc_2020}. % , higgins_kimball_2020}. 
Widespread acceptance of Trump's prolonged baseless insistence about the stolen election ultimately led to the storming of the US Capitol by Trump supporters on January 6, 2021~\cite{claire2022storm}. 
The intense competition between Trump and Biden started long before the election when parties had run presidential and congressional campaigns that cost almost \$14 billion---more than double the amount for the 2016 election \cite{horton_glatte_auer_2020}. %schwartz_2020, 

The advent of technology has reshaped the landscape of election campaigns; politicians across the world actively use social media to communicate with a general audience~\cite{hong2013benefits}. %hong2011does, 
Trump's Twitter account, the @realdonaldtrump handle, had 88.7 million followers at the time it was suspended in January 2021. It was one of the major channels Trump used to communicate with the public during his campaign. Communications generated on social media about US Election 2020 and its related topics had provided us with a large volume of data for this study.

\section{Related Work}

\subsection{Opinion Mining on Social Media}

Twitter is one of the most popular social media, which began as an SMS text-based service. 
The micro-blog platform originally has a length limit of 140 characters for each tweet, which was driven by the 160-character length of SMS. Over time, Twitter evolved to allow up to 280 characters for each tweet. 
As opposed to traditional media, Twitter has provided a platform for users to produce media content and voice their opinion publicly. This spawned the creation of a large public collection of opinions for many events happening around the globe~\cite{kwak2010twitter, wang2012system, conover2011predicting}.

Many fields have enjoyed rapid growth since the creation of social media, as it is the first time in human history that researchers get access to a huge volume of digital data expressing individual opinions. Opinion mining, the field of studying and analyzing people's opinions and sentiments towards an entity, has certainly been one of them~\cite{chen2010ai,fang2015sentiment,kharde2016sentiment}.
\citet{chen2010ai} had applied opinion mining on online forum messages to understand business constituents' opinions towards a company. 
% \citet{kumar2017systematic} discussed about the application of opinion mining on big data in areas such as government intelligence and market intelligence. 
\citet{fang2015sentiment} and \citet{kharde2016sentiment} leveraged the huge volume of data on Twitter to conduct studies on sentimental analysis.

\subsection{Stance Detection}

Stance detection is the task of automatically determining from the text whether the author of the text is in favor of, against, or neutral towards a proposition or target~\citep{mohammad2016semeval}. Stance is usually expressed in the form of categorical values, such as ``\textbf{Support}", ``\textbf{Neutral}" and ``\textbf{Against}". Stance detection can be applied in a variety of downstream tasks, for example, to classify the stance of posts on online debate forum~\cite{walker2012stance} 
% , hasan2013stance, sridhar2014collective}, 
and to help detect fake news and false rumours \cite{ma2018detect}.
% zubiaga2016analysing,ma2018detect,mendoza2010twitter,yang2022weakly}. 

%It has been shown that rumours that are later proven to be false tend to spark significantly larger numbers of denying tweets than rumours that are later confirmed to be true \citep{zubiaga2016analysing, mendoza2010twitter}.

Stance detection can be divided into two categories \cite{ghosh2019stance}. \textbf{Target-specific stance detection} considers each individual target separately,  whereas \textbf{multiple-target stance detection} jointly detects stance towards multiple related targets \cite{sobhani2017dataset}. 
In this study, a target-specific stance detection model is built for each of the three topics to classify \emph{stance of individual users}.

There are two ways to approach a stance detection task. Unsupervised learning involves capturing user clusters and labeling the stance for each cluster \cite{darwish2020unsupervised}. While it is effective at identifying clusters of vocal users with similar stances, it fails on less vocal users who only post very few tweets~\cite{samih2020few}. 
On the other hand, supervised learning leverages labeled training data to perform classification. Many supervised classification methods are developed for this purpose~\cite{igarashi2016tohoku, lai2020multilingual}.
However, it requires obtaining a large amount of training data, which can be expensive. 

%In the past, researchers trained machine learning models using linguistic and structural features extract manually \cite{somasundaran2010recognizing, sridhar2015joint}. \citet{dey2017twitter} proposed a feature-driven two-phase Support Vector Machine (SVM) architecture, \citet{hacohen2017stance} used 18 features including neural character skip-ngrams and character ngrams with 8 variants of machine learning classification models. \citet{sen2018stance} proposed a novel set of features involving a SVM model and feed-forward neural network model. \citet{kuccuk2018stance} used unigrams, bigrams, hashtags, external links, emoticons, and name entities as features to an SVM models. Neural models were developed for stance detection tasks in recent years and had obtained better results. \citet{du2017stance} created a novel bidirectional LSTM-based attention-based model for stance classification. \citet{zarrella2016mitre}, the winning team of SemEval 2016 Task 6A challenge, proposed a transfer learning method with features learned via distant supervision on two large unlabeled datasets. \citet{wei2016pkudblab}, the second position holders of SemEval 2016 Task 6A challenge, used a CNN-based sentence classification model \cite{DBLP:journals/corr/Kim14f}. \citet{chen2016utcnn} applied neural network model to classify stance of social media posts by considering user taste, topic taste and user comments on posts.

\subsection{Language Models}

Since the development of Recurrent Neural Networks (RNN) and Long Short-Term Memory (LSTM), they have been used widely in the area of language modeling and machine translation.
However, RNN and LSTM are unable to enjoy the full benefit of the huge computation power we have now with GPU and parallelization due to their sequential computation design. The attention mechanism provides an effective way for sequence modeling by leveraging parallelization, allowing the model to capture dependencies regardless of their distance in the sequence. 
In 2017, Google proposed the Transformer model \cite{vaswani2017attention} for sequence transduction tasks. BERT (Bidirectional Encoder Representations from Transformer) based on Transformers revolutionized the NLP world by producing state-of-the-art performance on numerous tasks~\cite{zhang2020semantics, wang2019multi}. It is pre-trained on a large corpus of unlabeled text and can be fine-tuned with just one additional output layer to perform a wide range of tasks.

\subsection{Distant Supervision}

As deep learning requires a large amount of labeled training data, manual labeling becomes more and more costly and non-scalable. On social media, distant supervision using  hashtags has been widely used to obtain weak labels for political affiliation~\cite{maulana2020measuring}, the emotion of tweets~\cite{hasan2014using}, or stance toward constitutional referendum~\cite{di2021content} due to its cost-effectiveness. 
We propose a more robust hashtag-based labeling method for a stance detection task on a huge corpus of English tweets. 
Instead of discarding tweets containing hashtags of opposing stances, we design a three-way classifier with a non-opinionated class. It classifies tweets based on a scoring approach. A neural stance detection model based on BERT is subsequently built in the next step to achieve even higher performance on a small sample of ground truth.

\section{Our Dataset}

Since the event, US Election 2020, is closely related to mask adoption and racial unrest events, we chose Trump as the target topic of the connected behavior study and selected wearing masks and racial equality as related topics to study their impact on people's behavior during the US Election 2020. 

The selection criteria for wearing masks and racial equality were based on four main factors. Firstly, both of them are highly controversial events which had spurred huge debates on social media and created a large volume of content containing individual opinions to support our study. Secondly, COVID-19 response measures and racial equality were among the issues covered in the presidential debate \cite{kaplan_michael_2020}. The clear ideological division between Trump and Biden could induce political partisanship in people's behaviors. 
% \cite{milosh2020political}. 
Next, topics should follow a chronological order for us to model causal effect, i.e., the events of mask adoption and racial unrest should take place before the US Election to affect people's decisions in the election, hence validating the ability to predict human behavior. Lastly, both topics were selected for taking place in the same year as US Election 2020 to minimize opinion drift over time.

% Since we aim to understand and uncover connected behavior among the three controversial issues: wearing masks, racial equality, and US election, we need to detect the stances of users on those three topics. In doing so, we collect tweets related to multiple selected topics. The challenge we faced was that in order to understand connected behavior of an user on several topics, the user need to have expressed an opinion towards all selected topics. As the number of topic grows, the pool of qualified users gets smaller. As such, a large number of tweets need to be collected to get a sizable number of users.

% \subsection{Data Collection}
\subsection{Topic-related Twitter Data}

For this study, we collected a total of 106,580,239 tweets on three selected events from Twitter using its official Academic API~\cite{twitteracademicapi}.
For each topic, we collected tweets using topic-related keywords in the API query. We carefully selected those keywords via a systematic approach. We first queried 100 tweets for each topic using root keywords describing the topic, such as `election' for the topic Trump, `COVID' and `mask' for the wearing masks topic, and `George Floyd' and `\#blacklivesmatter' for racial equality topic~\cite{yang2016narrative}. Then, we extracted a list of topic-related keywords by inspecting those tweets. The keywords selected were mostly generic words, phrases, or hashtags related to each topic, to avoid introducing stance bias in the data collection process. 
Full lists of the keywords used for data collection are listed in Appendix~\ref{app:data_collection_keywords}.

Specifically, we collected 81,028,363 Trump-related tweets for a period of one month between November 3 to December 2, 2020. Trump-related Twitter activity for this period was extremely high as people were very expressive about their political inclination.
We collected 14,834,164 wearing masks tweets from April 3 to November 2, 2020. CDC made a statement on April 3, 2020, recommending people to wear face masks following a record high daily death due to COVID-19. It has spurred a debate about mask adoption, and many people voiced their opinion on Twitter. 
For racial equality, a few controversial events happened in 2020, including the tragic incident of George Floyd, which sparked a huge debate on social media platforms~\cite{barkan2020high,weine2020justice}. We gathered 10,717,712 racial equality related tweets from May 25, 2020, the date George Floyd's incident took place, to November 2, 2020. 
Data collection time range for both wearing masks and racial equality topics were selected to benefit from the explosion of Twitter activity due to major events and to ensure tweets collected were close to the US Election 2020 to minimize opinion drift. 
% In total, we collected 106,580,052 tweets across the three topics. 

\subsection{Data Cleaning}

% Raw data is extracted by Twitter API using keywords and time based filters. 
% However, tweets collected using keywords and hashtags can be very noisy, they may contain bots, spammers and unreliable users. 
To further clean our dataset, we removed retweets, bot-likely accounts, and less active accounts to retain users of interest.  
% We performed data cleaning to remove all unwanted tweets and only retained users of interest. 
% There are approximately 29 million tweets for further analysis after applying data cleaning techniques described in subsequent sections.

% \subsubsection{Retweets}
% \paragraph{Retweets.}
\noindent \textbf{Retweets.}
For this study, we only look at original tweets to detect users' stance towards target topics as retweets or quoted tweets may not always mean a user endorsing the stance of the original tweets~\cite{opgenhaffen2014social}, which could add noise if included. Thus, we excluded all retweets and quoted tweets from our dataset. 
% boyd2010tweet,

% Retweet is a feature on Twitter for user to re-post another person's tweet to all his followers. There are many reasons on why a person would retweet another user's tweet, they may not always mean the user endorses the stance of the original user~\cite{boyd2010tweet,opgenhaffen2014social}. 
% In this study, we only look at original tweets to detect users' stance towards target topics. 
% Although Twitter API allows researchers to specify whether tweets posted using retweet function on Twitter should be returned from a query, some tweets were retweeted by adding an "RT" before the original tweet, bypassing Twitter's retweet function. Hence, a filter is applied to remove all retweets.

% \subsubsection{Bots and Spammer}
% \paragraph{Bots and Spammer.} 
\noindent \textbf{Bots and Spammer.} 
We further removed bot-likely accounts to better capture individual opinions of real human users as they tend to be automated to post information~\cite{ferrara2016rise}.
% kantepe2017preprocessing, subrahmanian2016darpa, chavoshi2016debot, knauth2019language
% Twitter has many accounts that are controlled by computer programs, which are known as bots. Most bots were created to gain followers, support sponsored activities, sell products, and to post information to influence community on a specific topic \cite{chavoshi2016debot,kantepe2017preprocessing,ferrara2016rise,subrahmanian2016darpa,knauth2019language}.
% According to a Twitter SEC filing, approximately 8.5\% of Twitter users are bots \cite{linshi_2014}. 
One of the strong signatures of these bot accounts is that they leave a much larger amount of tweets than genuine users typically do.
We calculated the number of posts per user and removed the top 5 percentile by user post count as spammer accounts.
We manually examined those highly active users with different thresholds for bots removal and observed a significantly higher percentage of tweets having no opinion at the top 5 percentile of users.

% \subsubsection{Unreliable User}
% \paragraph{Unreliable User.} 
\noindent \textbf{Users who rarely post.} 
% When a person has a stance towards a certain topic, that person is likely to talk about this topic and post on social media more often. In contrast, when a person only post about a topic once during the entire time frame, it might be due to that person unintentionally posted content which used key words we had defined for the topic. 
To more reliably detect one's stance towards a topic, we excluded those users who have only posted one tweet for the given topic in the defined time frame.
% Furthermore, the more tweets an user post about a topic, the more accurate we can detect his stance towards that topic. Hence, users who have only posted one topic related tweet in the defined time frame were removed due to its limited reliability.

% Table \ref{tab:cleaning_summary} summarised the result of applying above mentioned data cleaning techniques. There were about 29 million tweets across all three topics after cleaning.
The above three cleaning steps left us with 29 million tweets across the three topics. 

\subsection{User-level Twitter Data}
We further collected users' profile information (i.e., location, description, protected account, verified account, \# of followers, \# of friends, \# of tweets, and \# of lists that include the user) and their historical tweets using Twitter API as baseline information to validate the additional prediction power that stance can provide. 
% User's profile information and historical tweets were collected from Twitter as baseline information to validate the additional prediction power stance can contribute. 
% Profile information collected using Twitter API include 1) location specified in user's profile ; 2) text of user's profile description; 3) whether user has chosen to protect their Tweets; 4) whether user is verified by Twitter; 5) URL specified in user's profile if present; 6) number of users who follow this user; 7) number of users this user is following; 8) number of tweets (included retweets) posted by this user; 9) number of lists that include this user. 
For each user, we collected the most recent 100 tweets between November 3, 2019 to November 2, 2020 to engineer individual-level features.
% Each user's most recent 100 tweets between November 3, 2019 to November 2, 2020 were collected to engineer user's content feature.

\subsection{User Social Network}
In order to build a social network graph to validate the quality of stances obtained from the proposed framework, we collected retweets from the most active 10,000 users ranked by the number of posts for a period of 1 year prior to the US Election 2020.

% \begin{table}
% \footnotesize
% \begin{tabular}{|l|c|c|c|c|}
%  \hline
%  Topic & Date Range & Item & Before Cleaning & After Cleaning \\
%  \hline
%  \multirow{2}{3em}{Election} & \multirow{2}{5.5em}{3 Nov 2020 - 2 Dec 2020} 
%  & No. of Tweets & 81,028,363 & 18,784,352 \\ \cline{3-5}
%  && No. of Users & 7,674,046 & 2,013,659 \\ 
%  \hline
%  \multirow{2}{3em}{Mask} & \multirow{2}{5.5em}{3 Apr 2020 - 2 Nov 2020} 
%  & No. of Tweets & 14,834,164 & 2,871,852  \\ \cline{3-5}
%  && No. of Users & 4,967,681 & 617,412  \\ 
%  \hline
%  \multirow{2}{3em}{Racial} & \multirow{2}{5.5em}{25 May 2020 - 2 Nov 2020}  
%  & No. of Tweets & 10,717,712 & 7,352,555  \\ \cline{3-5}
%  && No. of Users & 243,806 & 223,471  \\ 
%  \hline
% \end{tabular}
% \caption{Data cleaning summary}
% \label{tab:cleaning_summary}
% \end{table}

\section{Distantly Supervised Stance Detection}

Supervised learning requires labeled training data. However, manually labeling a large amount of data is an expensive and time-consuming task \cite{chen2016statistical, magdy2015distant}. %deriu2017leveraging,
Many previous works on NLP classification tasks obtained manually annotated data via crowdsourcing \cite{becker2011beyond},  %
% , ghenai2017catching}, , sparks2016facility
but it was costly and sometimes the results were unreliable if the workers did not have sufficient knowledge of the topic of tweets they were assigned to annotate. To overcome these challenges, we developed a hashtag-based stance labeling method to obtain weak labels for millions of tweets with little human effort. Then, we use distant supervision techniques to build a BERT-based stance detector.

\begin{table}[t]
\centering
\caption{Result of hashtag-based stance labeling.}
\small
\begin{tabular}{c|c|c|c}
 \hline
  & \textbf{Against} & \textbf{Non-opinionated} & \textbf{Support} \\
 \hline
 \hline
 Trump & 209,662 & 1,741,650 & 250,625 \\
 \hline
 Wearing Masks & 18,677 & 486,706 & 192,357 \\
 \hline
 Racial Equality& 46,763 & 1,648,350 & 253,579 \\
 \hline                           
\end{tabular}
\label{tab:number_tweets_with_hashtag_by_stance}
\end{table}

\begin{table*}[ht!]
\centering
\caption{Stance detection results. Performance is measured using macro F1 score rounded off to two decimal places. The best performing model for each topic with hashtag masked is shown in bold.}
\resizebox{\textwidth}{!}{\begin{tabular}{|l|l|l|c|c|c|c|c|c|c|}
 \hline
 \multirow{3}{4em}{Topic} & \multirow{3}{4em}{Sampling Method} & \multirow{3}{4em}{\# of Samples} & \multicolumn{3}{c|}{Baseline} & \multicolumn{4}{c|}{BERT} \\ \cline{4-10}
                        & & & \multirow{2}{3.5em}{Random} & \multirow{2}{1.5em}{NB} & \multirow{2}{2.5em}{SVM} & \multicolumn{2}{c|}{Train w/ hashtag} & \multicolumn{2}{c|}{Train w/ hashtag masked} \\ \cline{7-10}
                        & & & & & & Test w/ hashtag & Test w/o hashtag & Test w/ hashtag & Test w/o hashtag \\ 
 \hline
 \multirow{3}{4em}{Trump} & Random & 100,000 & 0.27 & 0.33 & 0.38 & 0.99 & 0.30 & 0.42 & 0.47\\ \cline{2-10}
                             & \multirow{2}{4em}{Stratified} & 54,000 & 0.33 & 0.55 & 0.52 & 0.99 & 0.19 & 0.57 & 0.58 \\ \cline{3-10} 
                                                                      & & 90,000 & 0.33 & 0.56 & 0.53 & 1.00 & 0.18 & 0.60 & \textbf{0.60} \\
 \hline
 \multirow{2}{4em}{Mask} & Random & 100,000 & 0.28 & 0.34 & 0.54 & 1.00 & 0.35 & 0.34 & 0.58\\ \cline{2-10}
                         & Stratified & 54,000 & 0.33 & 0.59 & 0.61 & 0.99 & 0.35 & 0.49 & \textbf{0.69}\\ 
 \hline
 \multirow{4}{4em}{Racial} & Random & 100,000 & 0.24 & 0.34 & 0.45 & 1.00 & 0.31 & 0.43 & 0.54 \\ \cline{2-10}
                           & \multirow{3}{4em}{Stratified} & 54,000 & 0.33 & 0.54 & 0.59 & 1.00 & 0.17 & 0.66 & 0.67 \\ \cline{3-10}
                                                           & & 90,000 & 0.33 & 0.63 & 0.60 & 1.00 & 0.17 & 0.70 & 0.68 \\ \cline{3-10}
                                                           & & 135,000 & 0.33 & 0.63 & 0.62 & 1.00 & 0.17 & 0.70 & \textbf{0.69} \\
 \hline
\end{tabular}}
\label{tab:stance_detection_model_result}
\end{table*}

\begin{table*}[ht!]
\centering
\caption{Model performance on the 100 tweets randomly sampled and manually labeled. Macro F1 score for best performing model are shown in bold.}
\resizebox{\textwidth}{!}{\begin{tabular}{|c|c|c|c|c|c|c|c|c|c|c|}
 \hline
  \multicolumn{2}{|c|}{} & \multicolumn{3}{c|}{Baseline} & \multicolumn{3}{c|}{Hashtag-based stance labeling} & \multicolumn{3}{c|}{BERT-based stance detector}\\
 \cline{3-11}
  \multicolumn{2}{|c|}{} & Random & NB & SVM & Against & Non-opinionated & Support & Against & Non-opinionated & Support \\
 \hline
 \multirow{3}{3em}{Gold Labels} & Against & - & - & - & 12 & 21 & 0 & 22 & 9 & 2 \\ \cline{2-11}
& Non-opinionated & - & - & - & 2 & 51 & 2 & 8 & 45 & 2 \\ \cline{2-11}
& Support & - & - & - & 0 & 9 & 3 & 1 & 3 & 8 \\
 \hline
 \multicolumn{2}{|c|}{F1 Score} & 0.29 & 0.56 & 0.51 & \multicolumn{3}{c|}{0.54} & \multicolumn{3}{c|}{\textbf{0.72}}\\
 \hline                           
\end{tabular}}
\label{tab:model_performance}
\end{table*}

\subsection{Hashtag-Based Stance Labeling}\label{sect:labeling}
% \section{Hashtag Based Stance Detector}

A hashtag is a word or phrase without space preceded by a hash symbol (\#), which is used as a keyword to indicate the content of a tweet or the topic it is related to~\cite{small2011hashtag}. 
%Hashtag was designed to create more effective platform searches and facilitate efficient information sharing \cite{simeon2015evaluating}. 
On Twitter, users primarily use hashtags to convey sentiments and opinions \cite{han2011lexical}. 
Many studies have shown the effectiveness of hashtags to be used in tweets analytic tasks \cite{ she2014tomoha,an2016greysanatomy}.

To label the stance of tweets using hashtags, a set of stance-expressing hashtags is required for both the ``Support'' and ``Against'' classes. To achieve this, we examined the top 50 most frequently used hashtags for each topic, and extracted a set of stance-expressing hashtags representing Support or Against for every topic. 
Some hashtags were used to reference a topic or attract attention, such as \#election2020.
These hashtags do not show any stance and were hence omitted from the study. 
Hashtags which were obvious in their stance, such as \#trumpisaloser, were added to their respective class of stance-expressing hashtags. 
For hashtags whose stance is not obvious to tell, we randomly sampled 20 tweets containing each of these hashtags to manually annotate them, before allocating them to their respective groups. 

It has been observed that users with a certain stance toward a topic tend to use similar hashtags. To improve the discriminating power of the hashtag-based stance labeling, we expanded the list of hashtags for the ``Support'' and ``Against'' classes by adding a strong stance-expressing hashtags discovered during the manual labeling process.

Another challenge with using hashtags to determine the stance of tweets is that a tweet can contain more than one hashtag, and tweets can even contain hashtags expressing opposing stances. 
For example, \#maga is a hashtag used mainly by Trump supporters to support Trump's election campaign. However, in the tweet \textit{"Yet again, \#MAGA \#Republicans @GOP refuse to hold \#TraitorTrump accountable for removal of @EPA protections. @realDonaldTrump Pile of poo \#TrumpIsALaughingStock \#TrumpCrimeFamilyForPrison \#TrumpIsACompleteFailure \#TrumpIsALoser \#ClimateEmergency \#ToxicTrump"}, the author used \#maga as a subject to be criticised. The tweet contains hashtags having both supporting (\#MAGA) and against (\#TraitorTrump, \#TrumpIsALaughingStock, \#TrumpCrimeFamilyForPrison, \#TrumpIsACompleteFailure, \#TrumpIsALoser, \#ToxicTrump) stance towards Trump. 
To address this issue, we assumed that, when a tweet contains more ``Support'' hashtags than ``Against'' hashtags, it is more likely to be a tweet with a ``Support'' stance and vice versa. With this, we created a hashtag-based scoring algorithm relying on the ratio between ``Support'' and ``Against'' hashtags for stance labeling. The detailed algorithm can be found in Appendix \ref{app: hashtag_stance_detector_algo}.
% for the detailed algorithm. 

By hashtag-based stance labeling, we produced stance labels for approximately 5 million tweets. Table~\ref{tab:number_tweets_with_hashtag_by_stance} presents the number of tweets for each stance. 
However, only about 15\% of all tweets contain hashtags. 
To further classify all 29 million tweets, we build a stance detection model, which is described in the next section. 
\vspace{-6pt}
% A stance detection model is required to classify stance for all 29 million tweets, which is described in the next subsection. 

\subsection{BERT-based Stance Detector}

%BERT is a very popular model widely used for language tasks. 
In a study that compares the performance of numerous language models for the stance detection task on two separate datasets, \citet{ghosh2019stance} observed that BERT outperforms other widely used models, such as convolution neural network and SVM. 
Similar to \cite{ghosh2019stance}, we used the pre-trained BERT-Base Uncased model with 12 layers, 768 hidden size, 12 self-attention heads, and 110M parameters in this study. 
The output of the first head of the final layer is fed to a feed-forward layer along with softmax, and the model is fine-tuned to perform a target-specific stance detection task. We split the data in 80\%:20\% ratio for training and testing, and report macro F1 score in Table~\ref{tab:stance_detection_model_result}.\vspace{6pt}

The BERT stance detection model was fine-tuned only on tweets labeled by the hashtag-based labeling method described in $\S$\ref{sect:labeling}. 
Since the majority of tweets did not contain a hashtag, we needed to ensure that the model tuned on tweets with hashtags can be generalized to tweets without hashtags. As such, we conducted experiments with various models trained on data with different sampling methods, training data sizes, and hashtag mask settings.

We allowed hashtags in both training and test data to be masked to examine the generalizability of the model.
%Due to constraints in both hardware and time, we could only sample a limited number of tweets to train the stance detection model. 
With randomly sampled 100,000 tweets, the model achieved excellent performance with F1 score of 0.99. However, this high performance was expected as stances were labeled based on hashtags.
The model's performance is only meaningful when hashtags in test data were masked. 
With hashtags masked in test data, the model's F1 score dropped significantly. It dropped to 0.3, 0.35, and 0.31 for Trump, wearing masks, and racial equality topics, respectively. 
One way to improve the generalizability of the model is to train it with masked hashtags so that the model learns semantic and linguistic features of input text rather than hashtags. 
After masking hashtags in training data, the performance increased to 0.47, 0.58, and 0.54 for the three topics, respectively. 

It is observed that the data is highly imbalanced. In Table~\ref{tab:number_tweets_with_hashtag_by_stance}, almost 80\% of tweets were labeled as ``Non-opinionated.'' Furthermore, only about 2\% of tweets expressed ``Against'' stance in wearing masks and racial equality data. This resulted in most stances being likely to be classified as the majority class. 
Thus, we used stratified sampling to address this issue. The stratified sampling method has effectively improved the model performance, achieving F1 score of 0.6 for Trump and about 0.68 for wearing masks and racial equality topics.

%Due to the limited number of tweets with minority stance for the topic mask adoption, we could only sample 54,000 tweets using stratified sampling method. 
From the results of Trump and racial equality topics, we found that increasing the size of training data does not yield a significant improvement in model performance.
As a result, we chose 54,000 tweets with masked hashtags by stratified sampling as our training data. 

We created three baseline models as references. We present their performance in Table~\ref{tab:stance_detection_model_result}. 
Both multinomial Naive Bayes and SVM classifiers were implemented using scikit-learn library~\cite{pedregosa2011scikit}. The SVM classifier was applied with a linear kernel and without class weight. CountVectorizer from scikit-learn library was used to remove English stop words and generate unigram features. As expected, BERT achieved better performance than baseline models in all sampling methods and sample size configurations.
%As expected, a model that randomly classifies stance for each tweet had the lowest performance. Multinomial Naive Bayes classifier and Support Vector classifier achieved similar performance, albeit lower than BERT in all sampling method and sample size configurations.

\begin{figure*}[t!]
    \centering
    \includegraphics[width=15cm]{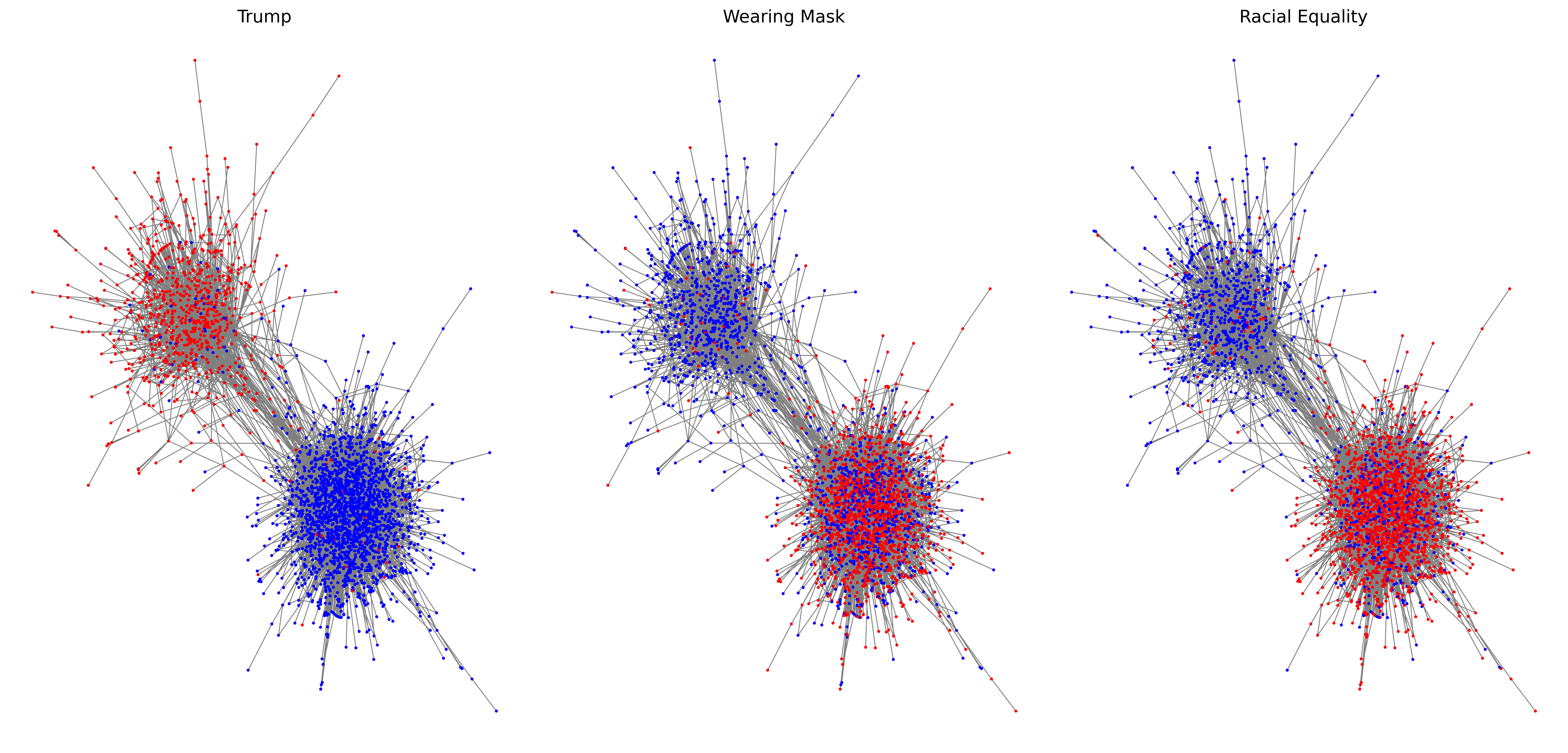}
    \caption{Retweet Networks of the three topics. A node represents a user, and an edge represents a retweet between two users. Nodes are colored by the user's stance towards each topic. Red denotes Support and blue denotes Against. All three networks show that the two clusters are well separated.}
    \label{fig:retweet_graph}
\end{figure*}

\subsection{Model Performance}

The BERT-based stance detection model was trained using labeled data from the hashtag-based stance labeling method.  It is important to analyze the performance of both models to understand the quality of data we have for studying connected behavior. To achieve this, we randomly sampled 100 tweets from Trump data and manually annotated them. These manually annotated stances are treated as ``gold labels'' to be compared with results from baseline models and two stance detectors proposed. Table~\ref{tab:model_performance} summarized the result.

Although the F1 score for hashtag-based stance labeling is only 0.54, this is largely driven by a low recall. Many ``Support'' and ``Against'' tweets were classified as ``Non-opinionated'' because limited hashtags were chosen for hashtag-based labeling to avoid false positives. Instead, the accuracy is high as shown in Table \ref{tab:model_performance}. Having high accuracy is crucial as it allows us to build a good BERT-based stance classifier based on quality data. 
The BERT-based stance detection was able to achieve a significantly better performance of 0.78, overcoming the limitation of having insufficient stance expressing hashtags. 
For example, the hashtag \#Trump is frequently used by both Trump supporters and haters, a tweet with \#Trump is labeled as Non-opinionated by the hashtag-based stance labeling method rather than either ``Support'' or ``Against''. In the tweet "\emph{Time to get serious people...we need to get \#Trump out of there...hes broken so many laws...}", the BERT-based model successfully captured the meaning of the entire sentence and classified it as ``Against'' Trump. 
%The outstanding performance of BERT-based stance detector has certainly boosted our confidence in the validity of connected behavior analysis.

\subsection{Validation Using A Retweet Network}

Retweet has been shown to be an effective network feature to reveal the stances of users~\cite{darwish2020unsupervised}.
To validate the quality of stances obtained using our proposed distantly supervised stance detection framework, we performed a network analysis on the most 10,000 active users ranked by the number of posts. The retweet network is created with nodes representing users and edges representing retweets between users. 
%\js{are these retweets new dataset? Then, may need more details (or potentially include them to the data collection section) } %Resolved
Nodes without any edges were dropped, resulting in a network with 3,104 nodes and 9,546 edges. In Figure~\ref{fig:retweet_graph}, users formed two clusters with dense connections within the cluster and sparse connections between them. The ratio of intra-stance edges to inter-stance edges is 4.26, 1.14, and 1.79 for topics of Trump, wearing masks, and racial equality, respectively. Coloring each node by stance obtained via the proposed framework, we observed that the majority of users within the same cluster have the same stance. 
%\js{Please make Figure 1 fit better to the paper (no white space), and also, I think the results should be presented numerically rather than visually. like confusion metrics or other measures.} %Resolved

\section{Connected Behavior Analysis}

Connected behavior analysis is the study of the relationship between people's behaviors in related events.
To examine connected behavior among the three topics, users who have shown a stance on all three topics are required. 
This reduced the size of the dataset from the result of the BERT-based stance classifier. 
A total of 183,848 Twitter users satisfied this condition.
Moreover, we found that the high proportion of Non-opinionated tweets overshadowed the number of tweets containing a stance. 
We thus removed all Non-opinionated tweets and only consider tweets with ``Support'' or ``Against'' stances. In total, we have 99,342 users who have expressed ``Support'' or ``Against'' towards all three topics.

\begin{figure}[h!]
    \centering
    \includegraphics[width=8cm]{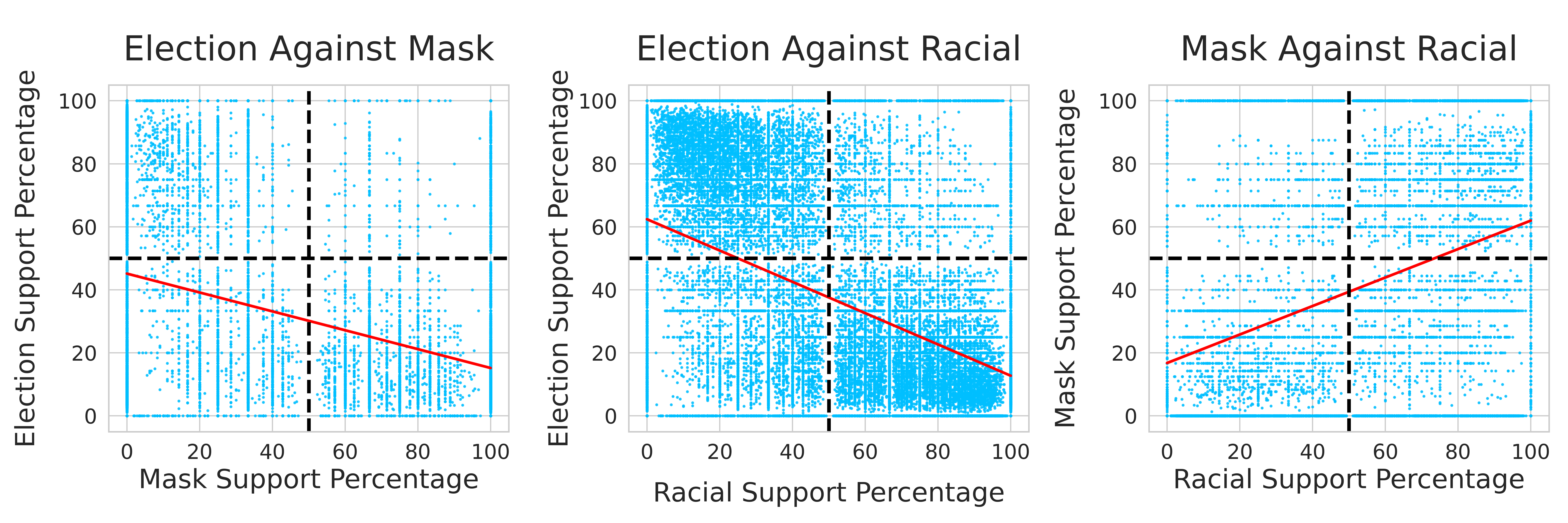}
    \caption{Correlation between stance of each topic pair}
    \label{fig:topic_correlation}
    \Description{Left and middle chart show dense clusters at top left and bottom right quadrant, and right chart shows dense clusters at bottom left and top right quadrant. Coupled with regression line, it demonstrated strong correlation between stance one each pair of topics.}
\end{figure}

Next, we aggregated stances at the user level by computing the percentage of tweets with a Support stance out of all topic-related tweets posted by a user. 
% One way of aggregating stances at the user level is to compute the percentage of tweets with a Support stance out of all topic-related tweets posted by a user. 
The percentage of Support shows the inclination of a user towards supporting that topic. 
Figure~\ref{fig:topic_correlation} illustrated a correlation of the percentages of Support between each pair of topics. Black dashed lines divide each subplot into 4 quadrants. 
% Each quadrant represented a permutation of stance towards two selected topics. It had shown a polarised result and a strong correlation between topics. 
Most people who support wearing masks were against Trump, most people who support racial equality were also against Trump, and most people who were against racial equality were against wearing masks. 
Hypothesis tests were performed to verify the significance of Spearman's rank correlation coefficient between each pair of topics. As shown in Table~\ref{tab:spearman_correlation}, Spearman's rank correlations between all topics were significant at $p<0.05$ significance level.

\begin{table}[h!]
%\small
\centering
\caption{Spearman's rank correlation coefficients and 2-tailed p-values}
\begin{tabular}{c|c|c|c}
 \hline
   \multicolumn{2}{c|}{Topics} & Correlation & p-value \\
 \hline
 \hline
 Trump & Wearing Masks & -0.34 & 0.0 \\
 \hline
 Trump & Racial Equality & -0.44 & 0.0 \\
 \hline
 Wearing Masks & Racial Equality & 0.32 & 0.0 \\
 \hline                           
\end{tabular}
\label{tab:spearman_correlation}
\end{table} 

% \begin{figure}[h]
%     \centering
%     \includegraphics[width=6cm]{Figures/prob_voting_trump_report_45_55.png}
%     \caption{Probability for voting Trump}
%     \label{fig:prob_voting_trump}
% \end{figure}

If users have similar percentages of Support and Against tweets for a single topic, their stances can be considered weak. We omit them for further analysis. By a manual inspection, we empirically set the percentage threshold as 55\%; 
when users have a percentage of Support greater than or equal to 55\%, they are likely to support the topic. However, when users have support percentage smaller or equal to 45\%, they are likely to be against the topic. 
Accordingly, we categorized every user as either ``Support'' or ``Against'' towards each of the three topics. 
The result has demonstrated the presence of strongly connected behavior between topics of wearing masks, racial equality, and Trump. 
As illustrated in Table~\ref{tab:prob_voting_trump}, if a person supports wearing masks and racial equality, there is only a probability of 6\% that they would vote for Trump. 
By contrast, if a person is against wearing masks and racial equality, there is a very high probability of 71\% that he would vote for Trump.

\begin{table}[h!]
%\small
\centering
\caption{Probability of a person voting for trump given his stance on wearing masks and racial equality.}
\begin{tabular}{l|l|c|c|}
    \multicolumn{2}{c}{}&\multicolumn{2}{c}{\textbf{Wearing Masks}}\\
    \cline{3-4}
    \multicolumn{2}{c|}{} & Support & Against \\
    \cline{2-4}
    \multirow{2}{*}{\textbf{Racial Equality}} & Support & 0.06 & 0.17 \\
    \cline{2-4}
    & Against & 0.19 & 0.71 \\
    \cline{2-4}
\end{tabular}
\label{tab:prob_voting_trump}
\end{table}

\section{Connected Behavior Prediction}
After establishing a strong correlation between people's stance on wearing masks, racial equality, and Trump, we performed an in-depth study to validate the capability of predicting a person's stance on Trump given his stances on wearing masks and racial equality. 
This analysis outlines the value of predicting people's behavior in a target event given their behaviors in related events and paves the way for further exploration.

\begin{table*}[ht!]
\centering
\caption{Ablation study of proposed features for predicting stance on Trump. Dim. indicates the dimension of features for each model.}
\resizebox{\textwidth}{!}{\begin{tabular}{|l|c|l|c|c|c|}
 \hline
 Model & \# & Features & Dim. & F1 & Accuracy \\
 \hline
 \multirow{2}{*}{Baseline} & 1 & Random & - & 0.47 & 0.50 \\
 & 2 & Majority Class & - & 0.42 & 0.73 \\
 \hline
 \multirow{3}{*}{Content-agnostic features} & 3 & Profile & 1,536 & 0.62 & 0.76 \\
 & 4 & Statistics & 7 & 0.48 & 0.73 \\
 & 5 & Prof. + Stats. & 1,543 & \textbf{0.64} & \textbf{0.77} \\
 \hline
 \multirow{5}{*}{Content Features} & 6 & Historical & 768 & 0.81 & 0.86 \\
 & 7 & Mask  & 768 & 0.76 & 0.82 \\
 & 8 & Racial  & 768 & 0.82 & 0.86 \\
 & 9 & Mask + Racial & 1,536 & 0.84 & 0.88 \\
 & 10 & Mask + Racial + Historical & 2,304 & \textbf{0.84} & \textbf{0.88} \\
 \hline
 Stance features & 11 & All & 2 & \textbf{0.80} & \textbf{0.84} \\
 \hline
 \multirow{7}{*}{Combination} & 12 & Row 5 + Mask Stance & 1,544 & 0.71 & 0.79 \\
 & 13 & Row 5 + Racial Stance & 1,544 & 0.76 & 0.82 \\
 & 14 & Row 5 + All Stance & 1,545 & 0.79 & 0.84 \\
 & 15 & Row 5 + Historical Cont. & 2311 & 0.81 & 0.86 \\
 & 16 & Row 15 + All Stance & 2,313 & 0.83 & 0.88 \\
 & 17 & Row 15 + Mask Cont. + Racial Cont. & 3,847 & 0.85 & 0.89 \\
 & 18 & Row 15 + Mask Cont. + Racial Cont. + All Stance & 3,849 & \textbf{0.85} & \textbf{0.89} \\
 \hline
\end{tabular}}
\label{fig:pred_model_results}
\end{table*}

\subsection{Features}

We use users' stances on wearing masks and racial equality as features to build a prediction model for their stance on Trump. 
In addition, we extract 1) content features from their historical tweets and 2) content-agnostic features from their profile and Twitter-related statistics to perform an ablation study.

\subsubsection{Content Features}
We used three content features for Trump stance prediction. 
We extracted mask and racial equality-related content features using their respective topic-related tweets. 
We also extracted historical content features using each user's most recent 100 tweets, excluding all topic-related tweets. 
% This is to ensure topic-related information is only contained in stance and topic-related content features. 
For each content feature, we encoded tweets with Sentence-BERT (SBERT)~\cite{reimers2019sentence} and compute their average embeddings as a user-level representation.

\subsubsection{Content Agnostic Features}
Twitter profile features were extracted from the profile description and URL listed in each user's profile. They were both encoded using SBERT and averaged to obtain user-level content-agnostic representation.
Twitter statistics features include: 1) number of followers, 2) number of followings, 3) number of tweets posted by the user, 4) number of public lists that the user is a member of, 5) number of days since the creation of account, 6) whether the account protects its tweets, and 7) whether the account is verified.

\subsection{Experiment Setup}
With 69,308 users, excluding deleted users whose profile or content could not be collected, we trained an XGBoost classifier to predict a user's stance on Trump. 
Stance features, content features, and content-agnostic features were evaluated separately and in combination to compare the model's performance. 
Numerical content-agnostic features were standardized, and categorical content-agnostic features and stance features were encoded as one-hot vectors in the pre-possessing stage. 
We split the data in an 80\%:20\% ratio for training and testing. Macro F1 score and accuracy were presented for each model. 

\subsection{Impact of Stance}
Table~\ref{fig:pred_model_results} shows the result of the connected behavior prediction model. 
Models using Twitter profile features, content features, and stance features exhibited substantially better performance than baseline models. 
Models with racial equality-related content features or stance consistently achieved better performance than models using those of wearing masks. It implies the racial equality topic is more closely related to Trump than wearing masks, making the stance on racial equality a better predictor for the stance on Trump. 
The model with only stance features on wearing masks and racial equality achieves an F1 score of 0.8, which highlights the importance of stance features and the potential of connected behavior prediction. 
Although the improvement is marginal, it is worth noting that content features of wearing masks and racial equality give a better performance than stance features. 
It is reasonable because those content features somehow encode stances regarding those topics as well, but they require a much larger model with remarkably higher dimensions. 
If we extend our model to cover more topics and issues, predictions based on stances only will have a huge advantage in terms of implementations over those based on all the content features. 

\section{Conclusion and Limitations}

In this study, we have collected a large volume of Twitter data on three social issues and proposed a hashtag-based mechanism to annotate training data with three-way stance labels. We also built a BERT-based distantly supervised model extending stance detection to all tweets collected regardless of hashtag inclusion. 
Connected behavior analysis has shown strong correlations between people's stances on wearing masks, racial equality, and Trump. 
Finally, as the first step towards connected behavior research, we studied the capability of predicting people's behavior in a target event (e.g., not voting for Trump) based on their behaviors in related events (e.g., wearing masks or support for racial equality). The result is promising. Stances toward related events are good predictors of the stance toward a target event.

% \section*{Limitations}

There are some limitations in this work. Firstly, the hashtag-based stance labeling method relies on a list of carefully curated hashtags, and its performance depends on the quality and quantity of stance-expressing hashtags picked. 
Due to various constraints, the 50 most frequently used hashtags for each topic were manually inspected for hashtag-based stance labeling. However, hashtags can exist in any words combinations, and 50 might be insufficient to cover the vast range of hashtags used by Twitter users. 
Hence, tweets containing stance-expressing hashtags that were not studied would have been classified as Non-opinionated using the hashtag-based stance labeling method, resulting in a low recall. 
The overall F1 score for the hashtag-based stance detector from the manual evaluation is 0.54, similar to that of machine learning baseline models. 
Having too few stance-expressing hashtags has led to imbalanced training data, and the quality of hashtags picked has a direct impact on labeling accuracy. 
If we examine more hashtags, we might expect an improvement in model performance. Alternatively, obtaining fully labeled data by humans would be ideal for training stance detection models, although being very costly. 
Secondly, bot removal was performed by evaluating users who have posted a significantly large amount of tweets within our dataset. A more effective method could be to use the BotOrNot service proposed by \citet{davis2016botornot}. The BotOrNot service uses more than 1,000 features, including network, user, temporal, content, and sentimental features, to classify an account as either a bot or human. The BotOrNot API has a rate limit of 180 requests per 15 minutes. Given enough time, the BotOrNot service could be applied to identify and remove bot accounts more accurately and allow us to focus on human users for connected behavior study.

% We are excited about the future of connected behavior study. We plan to extend the study of connected behaviors to more topics covering difference types of event. Making a better user stance model is another goal for our research.

\section*{Ethics Statement}
This work aims to predict human behaviors in target events given his behaviors in related events, a question that can be answered by analyzing individual-level Twitter data.
For individual Twitter accounts, we take several steps to limit the potential negative impacts of our work. We only collect publicly available tweets to detect the stance of users toward controversial events and report aggregate information. Considering the sensitive level of topics discussed, we refrain from singling out individuals.
Lastly, for sharing our tweet data, we will publish only a list of tweet IDs, without any text or account information, according to Twitter's guidelines.

\begin{acks}
    This research is supported by the Singapore Ministry of Education (MOE) Academic Research Fund (AcRF) Tier 1 grant and the SMU-A*STAR Joint Lab in Social and Human-Centered Computing (Grant No. SAJL-2022-CSS003).
\end{acks}

\bibliographystyle{unsrt}

%%% -*-BibTeX-*-
%%% Do NOT edit. File created by BibTeX with style
%%% ACM-Reference-Format-Journals [18-Jan-2012].

\begin{thebibliography}{64}

%%% ====================================================================
%%% NOTE TO THE USER: you can override these defaults by providing
%%% customized versions of any of these macros before the \bibliography
%%% command.  Each of them MUST provide its own final punctuation,
%%% except for \shownote{}, \showDOI{}, and \showURL{}.  The latter two
%%% do not use final punctuation, in order to avoid confusing it with
%%% the Web address.
%%%
%%% To suppress output of a particular field, define its macro to expand
%%% to an empty string, or better, \unskip, like this:
%%%
%%% \newcommand{\showDOI}[1]{\unskip}   % LaTeX syntax
%%%
%%% \def \showDOI #1{\unskip}           % plain TeX syntax
%%%
%%% ====================================================================

\ifx \showCODEN    \undefined \def \showCODEN     #1{\unskip}     \fi
\ifx \showDOI      \undefined \def \showDOI       #1{#1}\fi
\ifx \showISBNx    \undefined \def \showISBNx     #1{\unskip}     \fi
\ifx \showISBNxiii \undefined \def \showISBNxiii  #1{\unskip}     \fi
\ifx \showISSN     \undefined \def \showISSN      #1{\unskip}     \fi
\ifx \showLCCN     \undefined \def \showLCCN      #1{\unskip}     \fi
\ifx \shownote     \undefined \def \shownote      #1{#1}          \fi
\ifx \showarticletitle \undefined \def \showarticletitle #1{#1}   \fi
\ifx \showURL      \undefined \def \showURL       {\relax}        \fi
% The following commands are used for tagged output and should be
% invisible to TeX
\providecommand\bibfield[2]{#2}
\providecommand\bibinfo[2]{#2}
\providecommand\natexlab[1]{#1}
\providecommand\showeprint[2][]{arXiv:#2}

\bibitem[bbc(2020)]%
        {bbc_2020}
 \bibinfo{year}{2020}\natexlab{}.
\newblock \showarticletitle{US election: Trump says Biden won but again refuses
  to concede}.
\newblock \bibinfo{journal}{\emph{BBC}} (\bibinfo{date}{Nov}
  \bibinfo{year}{2020}).
\newblock
\urldef\tempurl%
\url{https://www.bbc.com/news/election-us-2020-54952098}
\showURL{%
\tempurl}


\bibitem[twi(2022)]%
        {twitteracademicapi}
 \bibinfo{year}{2022}\natexlab{}.
\newblock \bibinfo{title}{Twitter API for academic research | products |
  twitter developer platform}.
\newblock
\newblock
\urldef\tempurl%
\url{https://developer.twitter.com/en/products/twitter-api/academic-research}
\showURL{%
\tempurl}


\bibitem[Adolph et~al\mbox{.}(2022)]%
        {adolph2022governor}
\bibfield{author}{\bibinfo{person}{Christopher Adolph}, \bibinfo{person}{Kenya
  Amano}, \bibinfo{person}{Bree Bang-Jensen}, \bibinfo{person}{Nancy Fullman},
  \bibinfo{person}{Beatrice Magistro}, \bibinfo{person}{Grace Reinke}, {and}
  \bibinfo{person}{John Wilkerson}.} \bibinfo{year}{2022}\natexlab{}.
\newblock \showarticletitle{Governor partisanship explains the adoption of
  statewide mask mandates in response to COVID-19}.
\newblock \bibinfo{journal}{\emph{State Politics \& Policy Quarterly}}
  \bibinfo{volume}{22}, \bibinfo{number}{1} (\bibinfo{year}{2022}),
  \bibinfo{pages}{24--49}.
\newblock


\bibitem[An et~al\mbox{.}(2021)]%
        {an-etal-2021-predicting-anti}
\bibfield{author}{\bibinfo{person}{Jisun An}, \bibinfo{person}{Haewoon Kwak},
  \bibinfo{person}{Claire~Seungeun Lee}, \bibinfo{person}{Bogang Jun}, {and}
  \bibinfo{person}{Yong-Yeol Ahn}.} \bibinfo{year}{2021}\natexlab{}.
\newblock \showarticletitle{Predicting Anti-{A}sian Hateful Users on {T}witter
  during {COVID}-19}. In \bibinfo{booktitle}{\emph{Findings of the Association
  for Computational Linguistics: EMNLP 2021}}.
\newblock


\bibitem[An et~al\mbox{.}(2014)]%
        {an2014partisan}
\bibfield{author}{\bibinfo{person}{Jisun An}, \bibinfo{person}{Daniele
  Quercia}, {and} \bibinfo{person}{Jon Crowcroft}.}
  \bibinfo{year}{2014}\natexlab{}.
\newblock \showarticletitle{Partisan sharing: Facebook evidence and societal
  consequences}. In \bibinfo{booktitle}{\emph{Proceedings of the second ACM
  conference on Online social networks}}. \bibinfo{pages}{13--24}.
\newblock


\bibitem[An and Weber(2016)]%
        {an2016greysanatomy}
\bibfield{author}{\bibinfo{person}{Jisun An} {and} \bibinfo{person}{Ingmar
  Weber}.} \bibinfo{year}{2016}\natexlab{}.
\newblock \showarticletitle{\# greysanatomy vs.\# yankees: Demographics and
  Hashtag Use on Twitter}. In \bibinfo{booktitle}{\emph{Proceedings of the
  International AAAI Conference on Web and Social Media}},
  Vol.~\bibinfo{volume}{10}. \bibinfo{pages}{523--526}.
\newblock


\bibitem[Barber{\'a} et~al\mbox{.}(2015)]%
        {barbera2015tweeting}
\bibfield{author}{\bibinfo{person}{Pablo Barber{\'a}}, \bibinfo{person}{John~T
  Jost}, \bibinfo{person}{Jonathan Nagler}, \bibinfo{person}{Joshua~A Tucker},
  {and} \bibinfo{person}{Richard Bonneau}.} \bibinfo{year}{2015}\natexlab{}.
\newblock \showarticletitle{Tweeting from left to right: Is online political
  communication more than an echo chamber?}
\newblock \bibinfo{journal}{\emph{Psychological science}} \bibinfo{volume}{26},
  \bibinfo{number}{10} (\bibinfo{year}{2015}), \bibinfo{pages}{1531--1542}.
\newblock


\bibitem[Barkan(2020)]%
        {barkan2020high}
\bibfield{author}{\bibinfo{person}{Ross Barkan}.}
  \bibinfo{year}{2020}\natexlab{}.
\newblock \showarticletitle{High-profile Killings of Unarmed Black People Spark
  Calls for Reform}.
\newblock \bibinfo{journal}{\emph{ABAJ}}  \bibinfo{volume}{106}
  (\bibinfo{year}{2020}), \bibinfo{pages}{14--16}.
\newblock


\bibitem[Becker et~al\mbox{.}(2011)]%
        {becker2011beyond}
\bibfield{author}{\bibinfo{person}{Hila Becker}, \bibinfo{person}{Mor Naaman},
  {and} \bibinfo{person}{Luis Gravano}.} \bibinfo{year}{2011}\natexlab{}.
\newblock \showarticletitle{Beyond trending topics: Real-world event
  identification on twitter}. In \bibinfo{booktitle}{\emph{Proceedings of the
  International AAAI Conference on Web and Social Media}},
  Vol.~\bibinfo{volume}{5}. \bibinfo{pages}{438--441}.
\newblock


\bibitem[Bing et~al\mbox{.}(2014)]%
        {bing2014public}
\bibfield{author}{\bibinfo{person}{Li Bing}, \bibinfo{person}{Keith~CC Chan},
  {and} \bibinfo{person}{Carol Ou}.} \bibinfo{year}{2014}\natexlab{}.
\newblock \showarticletitle{Public sentiment analysis in Twitter data for
  prediction of a company's stock price movements}. In
  \bibinfo{booktitle}{\emph{2014 IEEE 11th International Conference on
  e-Business Engineering}}. IEEE, \bibinfo{pages}{232--239}.
\newblock


\bibitem[Chen et~al\mbox{.}(2016)]%
        {chen2016statistical}
\bibfield{author}{\bibinfo{person}{Chao Chen}, \bibinfo{person}{Yu Wang},
  \bibinfo{person}{Jun Zhang}, \bibinfo{person}{Yang Xiang},
  \bibinfo{person}{Wanlei Zhou}, {and} \bibinfo{person}{Geyong Min}.}
  \bibinfo{year}{2016}\natexlab{}.
\newblock \showarticletitle{Statistical features-based real-time detection of
  drifted Twitter spam}.
\newblock \bibinfo{journal}{\emph{IEEE Transactions on Information Forensics
  and Security}} \bibinfo{volume}{12}, \bibinfo{number}{4}
  (\bibinfo{year}{2016}), \bibinfo{pages}{914--925}.
\newblock


\bibitem[Chen and Zimbra(2010)]%
        {chen2010ai}
\bibfield{author}{\bibinfo{person}{Hsinchun Chen} {and} \bibinfo{person}{David
  Zimbra}.} \bibinfo{year}{2010}\natexlab{}.
\newblock \showarticletitle{AI and opinion mining}.
\newblock \bibinfo{journal}{\emph{IEEE Intelligent Systems}}
  \bibinfo{volume}{25}, \bibinfo{number}{3} (\bibinfo{year}{2010}),
  \bibinfo{pages}{74--80}.
\newblock


\bibitem[Cinelli et~al\mbox{.}(2021)]%
        {cinelli2021echo}
\bibfield{author}{\bibinfo{person}{Matteo Cinelli}, \bibinfo{person}{Gianmarco
  De~Francisci Morales}, \bibinfo{person}{Alessandro Galeazzi},
  \bibinfo{person}{Walter Quattrociocchi}, {and} \bibinfo{person}{Michele
  Starnini}.} \bibinfo{year}{2021}\natexlab{}.
\newblock \showarticletitle{The echo chamber effect on social media}.
\newblock \bibinfo{journal}{\emph{Proceedings of the National Academy of
  Sciences}} \bibinfo{volume}{118}, \bibinfo{number}{9} (\bibinfo{year}{2021}).
\newblock


\bibitem[Colleoni et~al\mbox{.}(2014)]%
        {colleoni2014echo}
\bibfield{author}{\bibinfo{person}{Elanor Colleoni},
  \bibinfo{person}{Alessandro Rozza}, {and} \bibinfo{person}{Adam Arvidsson}.}
  \bibinfo{year}{2014}\natexlab{}.
\newblock \showarticletitle{Echo chamber or public sphere? Predicting political
  orientation and measuring political homophily in Twitter using big data}.
\newblock \bibinfo{journal}{\emph{Journal of communication}}
  \bibinfo{volume}{64}, \bibinfo{number}{2} (\bibinfo{year}{2014}),
  \bibinfo{pages}{317--332}.
\newblock


\bibitem[Conover et~al\mbox{.}(2011)]%
        {conover2011predicting}
\bibfield{author}{\bibinfo{person}{Michael~D Conover}, \bibinfo{person}{Bruno
  Gon{\c{c}}alves}, \bibinfo{person}{Jacob Ratkiewicz},
  \bibinfo{person}{Alessandro Flammini}, {and} \bibinfo{person}{Filippo
  Menczer}.} \bibinfo{year}{2011}\natexlab{}.
\newblock \showarticletitle{Predicting the political alignment of twitter
  users}. In \bibinfo{booktitle}{\emph{2011 IEEE third international conference
  on privacy, security, risk and trust and 2011 IEEE third international
  conference on social computing}}. IEEE, \bibinfo{pages}{192--199}.
\newblock


\bibitem[Darwish et~al\mbox{.}(2020)]%
        {darwish2020unsupervised}
\bibfield{author}{\bibinfo{person}{Kareem Darwish}, \bibinfo{person}{Peter
  Stefanov}, \bibinfo{person}{Micha{\"e}l Aupetit}, {and}
  \bibinfo{person}{Preslav Nakov}.} \bibinfo{year}{2020}\natexlab{}.
\newblock \showarticletitle{Unsupervised user stance detection on twitter}. In
  \bibinfo{booktitle}{\emph{Proceedings of the International AAAI Conference on
  Web and Social Media}}, Vol.~\bibinfo{volume}{14}. \bibinfo{pages}{141--152}.
\newblock


\bibitem[Davis et~al\mbox{.}(2016)]%
        {davis2016botornot}
\bibfield{author}{\bibinfo{person}{Clayton~Allen Davis}, \bibinfo{person}{Onur
  Varol}, \bibinfo{person}{Emilio Ferrara}, \bibinfo{person}{Alessandro
  Flammini}, {and} \bibinfo{person}{Filippo Menczer}.}
  \bibinfo{year}{2016}\natexlab{}.
\newblock \showarticletitle{Botornot: A system to evaluate social bots}. In
  \bibinfo{booktitle}{\emph{Proceedings of the 25th international conference
  companion on world wide web}}. \bibinfo{pages}{273--274}.
\newblock
\urldef\tempurl%
\url{https://doi.org/10.1145/2872518.2889302}
\showDOI{\tempurl}


\bibitem[Del~Vicario et~al\mbox{.}(2016)]%
        {del2016spreading}
\bibfield{author}{\bibinfo{person}{Michela Del~Vicario},
  \bibinfo{person}{Alessandro Bessi}, \bibinfo{person}{Fabiana Zollo},
  \bibinfo{person}{Fabio Petroni}, \bibinfo{person}{Antonio Scala},
  \bibinfo{person}{Guido Caldarelli}, \bibinfo{person}{H~Eugene Stanley}, {and}
  \bibinfo{person}{Walter Quattrociocchi}.} \bibinfo{year}{2016}\natexlab{}.
\newblock \showarticletitle{The spreading of misinformation online}.
\newblock \bibinfo{journal}{\emph{Proceedings of the National Academy of
  Sciences}} \bibinfo{volume}{113}, \bibinfo{number}{3} (\bibinfo{year}{2016}),
  \bibinfo{pages}{554--559}.
\newblock


\bibitem[Desilver(2021)]%
        {desilver_2021}
\bibfield{author}{\bibinfo{person}{Drew Desilver}.}
  \bibinfo{year}{2021}\natexlab{}.
\newblock \showarticletitle{Turnout soared in 2020 as nearly two-thirds of
  eligible U.S. voters cast ballots for president}.
\newblock \bibinfo{journal}{\emph{Pew Research Center}} (\bibinfo{date}{Jan}
  \bibinfo{year}{2021}).
\newblock
\urldef\tempurl%
\url{https://www.pewresearch.org/fact-tank/2021/01/28/turnout-soared-in-2020-as-nearly-two-thirds-of-eligible-u-s-voters-cast-ballots-for-president/}
\showURL{%
\tempurl}


\bibitem[Di~Giovanni and Brambilla(2021)]%
        {di2021content}
\bibfield{author}{\bibinfo{person}{Marco Di~Giovanni} {and}
  \bibinfo{person}{Marco Brambilla}.} \bibinfo{year}{2021}\natexlab{}.
\newblock \showarticletitle{Content-based Stance Classification of Tweets about
  the 2020 Italian Constitutional Referendum}. In
  \bibinfo{booktitle}{\emph{Proceedings of the Ninth International Workshop on
  Natural Language Processing for Social Media}}. \bibinfo{pages}{14--23}.
\newblock


\bibitem[Fang and Zhan(2015)]%
        {fang2015sentiment}
\bibfield{author}{\bibinfo{person}{Xing Fang} {and} \bibinfo{person}{Justin
  Zhan}.} \bibinfo{year}{2015}\natexlab{}.
\newblock \showarticletitle{Sentiment analysis using product review data}.
\newblock \bibinfo{journal}{\emph{Journal of Big Data}} \bibinfo{volume}{2},
  \bibinfo{number}{1} (\bibinfo{year}{2015}), \bibinfo{pages}{1--14}.
\newblock


\bibitem[Ferrara et~al\mbox{.}(2016)]%
        {ferrara2016rise}
\bibfield{author}{\bibinfo{person}{Emilio Ferrara}, \bibinfo{person}{Onur
  Varol}, \bibinfo{person}{Clayton Davis}, \bibinfo{person}{Filippo Menczer},
  {and} \bibinfo{person}{Alessandro Flammini}.}
  \bibinfo{year}{2016}\natexlab{}.
\newblock \showarticletitle{The rise of social bots}.
\newblock \bibinfo{journal}{\emph{Commun. ACM}} \bibinfo{volume}{59},
  \bibinfo{number}{7} (\bibinfo{year}{2016}), \bibinfo{pages}{96--104}.
\newblock


\bibitem[Ghosh et~al\mbox{.}(2019)]%
        {ghosh2019stance}
\bibfield{author}{\bibinfo{person}{Shalmoli Ghosh}, \bibinfo{person}{Prajwal
  Singhania}, \bibinfo{person}{Siddharth Singh}, \bibinfo{person}{Koustav
  Rudra}, {and} \bibinfo{person}{Saptarshi Ghosh}.}
  \bibinfo{year}{2019}\natexlab{}.
\newblock \showarticletitle{Stance detection in web and social media: a
  comparative study}. In \bibinfo{booktitle}{\emph{International Conference of
  the Cross-Language Evaluation Forum for European Languages}}. Springer,
  \bibinfo{pages}{75--87}.
\newblock


\bibitem[Han and Baldwin(2011)]%
        {han2011lexical}
\bibfield{author}{\bibinfo{person}{Bo Han} {and} \bibinfo{person}{Timothy
  Baldwin}.} \bibinfo{year}{2011}\natexlab{}.
\newblock \showarticletitle{Lexical normalisation of short text messages: Makn
  sens a\# twitter}. In \bibinfo{booktitle}{\emph{Proceedings of the 49th
  annual meeting of the association for computational linguistics: Human
  language technologies}}. \bibinfo{pages}{368--378}.
\newblock


\bibitem[Hasan et~al\mbox{.}(2014)]%
        {hasan2014using}
\bibfield{author}{\bibinfo{person}{Maryam Hasan}, \bibinfo{person}{Emmanuel
  Agu}, {and} \bibinfo{person}{Elke Rundensteiner}.}
  \bibinfo{year}{2014}\natexlab{}.
\newblock \showarticletitle{Using hashtags as labels for supervised learning of
  emotions in twitter messages}. In \bibinfo{booktitle}{\emph{Acm sigkdd
  workshop on health informatics, new york, usa}}.
\newblock


\bibitem[Hong(2013)]%
        {hong2013benefits}
\bibfield{author}{\bibinfo{person}{Sounman Hong}.}
  \bibinfo{year}{2013}\natexlab{}.
\newblock \showarticletitle{Who benefits from Twitter? Social media and
  political competition in the US House of Representatives}.
\newblock \bibinfo{journal}{\emph{Government Information Quarterly}}
  \bibinfo{volume}{30}, \bibinfo{number}{4} (\bibinfo{year}{2013}),
  \bibinfo{pages}{464--472}.
\newblock


\bibitem[Horton et~al\mbox{.}(2020)]%
        {horton_glatte_auer_2020}
\bibfield{author}{\bibinfo{person}{Jake Horton}, \bibinfo{person}{Sarah
  Glatte}, {and} \bibinfo{person}{Soraya Auer}.}
  \bibinfo{year}{2020}\natexlab{}.
\newblock \showarticletitle{US election 2020: How much did it cost and who paid
  for it?}
\newblock \bibinfo{journal}{\emph{BBC}} (\bibinfo{date}{Nov}
  \bibinfo{year}{2020}).
\newblock
\urldef\tempurl%
\url{https://www.bbc.com/news/av/election-us-2020-54696386}
\showURL{%
\tempurl}


\bibitem[Igarashi et~al\mbox{.}(2016)]%
        {igarashi2016tohoku}
\bibfield{author}{\bibinfo{person}{Yuki Igarashi}, \bibinfo{person}{Hiroya
  Komatsu}, \bibinfo{person}{Sosuke Kobayashi}, \bibinfo{person}{Naoaki
  Okazaki}, {and} \bibinfo{person}{Kentaro Inui}.}
  \bibinfo{year}{2016}\natexlab{}.
\newblock \showarticletitle{Tohoku at SemEval-2016 task 6: Feature-based model
  versus convolutional neural network for stance detection}. In
  \bibinfo{booktitle}{\emph{Proceedings of the 10th International Workshop on
  Semantic Evaluation (SemEval-2016)}}. \bibinfo{pages}{401--407}.
\newblock


\bibitem[Kahane(2021)]%
        {kahane2021politicizing}
\bibfield{author}{\bibinfo{person}{Leo~H Kahane}.}
  \bibinfo{year}{2021}\natexlab{}.
\newblock \showarticletitle{Politicizing the mask: Political, economic and
  demographic factors affecting mask wearing behavior in the USA}.
\newblock \bibinfo{journal}{\emph{Eastern economic journal}}
  \bibinfo{volume}{47}, \bibinfo{number}{2} (\bibinfo{year}{2021}),
  \bibinfo{pages}{163--183}.
\newblock


\bibitem[Kallus(2014)]%
        {kallus2014predicting}
\bibfield{author}{\bibinfo{person}{Nathan Kallus}.}
  \bibinfo{year}{2014}\natexlab{}.
\newblock \showarticletitle{Predicting crowd behavior with big public data}. In
  \bibinfo{booktitle}{\emph{Proceedings of the 23rd International Conference on
  World Wide Web}}. \bibinfo{pages}{625--630}.
\newblock


\bibitem[Kaplan and Michael(2020)]%
        {kaplan_michael_2020}
\bibfield{author}{\bibinfo{person}{Thomas Kaplan} {and}
  \bibinfo{person}{D.~Shear Michael}.} \bibinfo{year}{2020}\natexlab{}.
\newblock \bibinfo{title}{Biden and Trump will debate six main issues. here's
  where they stand.}
\newblock
\newblock
\urldef\tempurl%
\url{https://www.nytimes.com/2020/09/29/us/politics/biden-on-the-issues-trump-on-the-issues-debate.html}
\showURL{%
\tempurl}


\bibitem[Kharde et~al\mbox{.}(2016)]%
        {kharde2016sentiment}
\bibfield{author}{\bibinfo{person}{Vishal Kharde}, \bibinfo{person}{Prof
  Sonawane}, {et~al\mbox{.}}} \bibinfo{year}{2016}\natexlab{}.
\newblock \showarticletitle{Sentiment analysis of twitter data: a survey of
  techniques}.
\newblock \bibinfo{journal}{\emph{arXiv preprint arXiv:1601.06971}}
  (\bibinfo{year}{2016}).
\newblock


\bibitem[Kwak et~al\mbox{.}(2010)]%
        {kwak2010twitter}
\bibfield{author}{\bibinfo{person}{Haewoon Kwak}, \bibinfo{person}{Changhyun
  Lee}, \bibinfo{person}{Hosung Park}, {and} \bibinfo{person}{Sue Moon}.}
  \bibinfo{year}{2010}\natexlab{}.
\newblock \showarticletitle{What is Twitter, a social network or a news
  media?}. In \bibinfo{booktitle}{\emph{Proceedings of the 19th international
  conference on World wide web}}. \bibinfo{pages}{591--600}.
\newblock


\bibitem[Lai et~al\mbox{.}(2020)]%
        {lai2020multilingual}
\bibfield{author}{\bibinfo{person}{Mirko Lai},
  \bibinfo{person}{Alessandra~Teresa Cignarella}, \bibinfo{person}{Delia
  Iraz{\'u}~Hern{\'a}ndez Far{\'\i}as}, \bibinfo{person}{Cristina Bosco},
  \bibinfo{person}{Viviana Patti}, {and} \bibinfo{person}{Paolo Rosso}.}
  \bibinfo{year}{2020}\natexlab{}.
\newblock \showarticletitle{Multilingual stance detection in social media
  political debates}.
\newblock \bibinfo{journal}{\emph{Computer Speech \& Language}}
  \bibinfo{volume}{63} (\bibinfo{year}{2020}), \bibinfo{pages}{101075}.
\newblock


\bibitem[Lee et~al\mbox{.}(2022a)]%
        {lee2022storm}
\bibfield{author}{\bibinfo{person}{Claire~Seungeun Lee}, \bibinfo{person}{Juan
  Merizalde}, \bibinfo{person}{John~D Colautti}, \bibinfo{person}{Jisun An},
  {and} \bibinfo{person}{Haewoon Kwak}.} \bibinfo{year}{2022}\natexlab{a}.
\newblock \showarticletitle{Storm the Capitol: Linking Offline Political Speech
  and Online Twitter Extra-Representational Participation on QAnon and the
  January 6 Insurrection}.
\newblock \bibinfo{journal}{\emph{Frontiers in Sociology}}  \bibinfo{volume}{7}
  (\bibinfo{year}{2022}).
\newblock


\bibitem[Lee et~al\mbox{.}(2022b)]%
        {claire2022storm}
\bibfield{author}{\bibinfo{person}{Claire~Seungeun Lee}, \bibinfo{person}{Juan
  Merizalde}, \bibinfo{person}{John~D. Colautti}, \bibinfo{person}{Jisun An},
  {and} \bibinfo{person}{Haewoon Kwak}.} \bibinfo{year}{2022}\natexlab{b}.
\newblock \showarticletitle{Storm the Capitol: Linking Offline Political Speech
  and Online Twitter Extra-Representational Participation on QAnon and the
  January 6 Insurrection}.
\newblock \bibinfo{journal}{\emph{Frontiers in Sociology}}  \bibinfo{volume}{7}
  (\bibinfo{year}{2022}).
\newblock
\showISSN{2297-7775}
\urldef\tempurl%
\url{https://doi.org/10.3389/fsoc.2022.876070}
\showDOI{\tempurl}


\bibitem[Ma et~al\mbox{.}(2018)]%
        {ma2018detect}
\bibfield{author}{\bibinfo{person}{Jing Ma}, \bibinfo{person}{Wei Gao}, {and}
  \bibinfo{person}{Kam-Fai Wong}.} \bibinfo{year}{2018}\natexlab{}.
\newblock \showarticletitle{Detect rumor and stance jointly by neural
  multi-task learning}. In \bibinfo{booktitle}{\emph{Companion proceedings of
  the the web conference 2018}}. \bibinfo{pages}{585--593}.
\newblock


\bibitem[Magdy et~al\mbox{.}(2015)]%
        {magdy2015distant}
\bibfield{author}{\bibinfo{person}{Walid Magdy}, \bibinfo{person}{Hassan
  Sajjad}, \bibinfo{person}{Tarek El-Ganainy}, {and} \bibinfo{person}{Fabrizio
  Sebastiani}.} \bibinfo{year}{2015}\natexlab{}.
\newblock \showarticletitle{Distant supervision for tweet classification using
  youtube labels}. In \bibinfo{booktitle}{\emph{Proceedings of the
  International AAAI Conference on Web and Social Media}},
  Vol.~\bibinfo{volume}{9}. \bibinfo{pages}{638--641}.
\newblock


\bibitem[Mahmud et~al\mbox{.}(2016)]%
        {mahmud2016predicting}
\bibfield{author}{\bibinfo{person}{Jalal Mahmud}, \bibinfo{person}{Geli Fei},
  \bibinfo{person}{Anbang Xu}, \bibinfo{person}{Aditya Pal}, {and}
  \bibinfo{person}{Michelle Zhou}.} \bibinfo{year}{2016}\natexlab{}.
\newblock \showarticletitle{Predicting attitude and actions of twitter users}.
  In \bibinfo{booktitle}{\emph{Proceedings of the 21st International Conference
  on Intelligent User Interfaces}}. \bibinfo{pages}{2--6}.
\newblock


\bibitem[Maulana and Situngkir(2020)]%
        {maulana2020measuring}
\bibfield{author}{\bibinfo{person}{Ardian Maulana} {and} \bibinfo{person}{Hokky
  Situngkir}.} \bibinfo{year}{2020}\natexlab{}.
\newblock \showarticletitle{Measuring Media Partisanship during Election: The
  Case of 2019 Indonesia Election}.
\newblock  (\bibinfo{year}{2020}).
\newblock


\bibitem[McClain(2021)]%
        {mcclain2021trump}
\bibfield{author}{\bibinfo{person}{Paula~D. McClain}.}
  \bibinfo{year}{2021}\natexlab{}.
\newblock \showarticletitle{Trump and racial equality in America? No pretense
  at all!}
\newblock \bibinfo{journal}{\emph{Policy Studies}}  \bibinfo{volume}{42}
  (\bibinfo{year}{2021}), \bibinfo{pages}{491--508}.
\newblock


\bibitem[Mejova et~al\mbox{.}(2022)]%
        {mejova2022modeling}
\bibfield{author}{\bibinfo{person}{Yelena Mejova}, \bibinfo{person}{Jisun An},
  \bibinfo{person}{Gianmarco De~Francisci Morales}, {and}
  \bibinfo{person}{Haewoon Kwak}.} \bibinfo{year}{2022}\natexlab{}.
\newblock \showarticletitle{Modeling political activism around gun debate via
  social media}.
\newblock \bibinfo{journal}{\emph{Transactions on Social Computing}}
  \bibinfo{volume}{5}, \bibinfo{number}{1-4} (\bibinfo{year}{2022}),
  \bibinfo{pages}{1--28}.
\newblock


\bibitem[Mohammad et~al\mbox{.}(2016)]%
        {mohammad2016semeval}
\bibfield{author}{\bibinfo{person}{Saif Mohammad}, \bibinfo{person}{Svetlana
  Kiritchenko}, \bibinfo{person}{Parinaz Sobhani}, \bibinfo{person}{Xiaodan
  Zhu}, {and} \bibinfo{person}{Colin Cherry}.} \bibinfo{year}{2016}\natexlab{}.
\newblock \showarticletitle{Semeval-2016 task 6: Detecting stance in tweets}.
  In \bibinfo{booktitle}{\emph{Proceedings of the 10th international workshop
  on semantic evaluation (SemEval-2016)}}. \bibinfo{pages}{31--41}.
\newblock


\bibitem[MONTANARO(2020)]%
        {domenico2020trump}
\bibfield{author}{\bibinfo{person}{DOMENICO MONTANARO}.}
  \bibinfo{year}{2020}\natexlab{}.
\newblock \showarticletitle{Majority Of Americans Say Trump Increased Racial
  Tensions After George Floyd's Death, Poll Finds}.
\newblock \bibinfo{journal}{\emph{The New York Times Company}}
  (\bibinfo{date}{Jun} \bibinfo{year}{2020}).
\newblock
\urldef\tempurl%
\url{https://www.npr.org/2020/06/06/871404645/majority-of-americans-say-trump-increased-racial-tensions-after-george-floyds-de}
\showURL{%
\tempurl}


\bibitem[O'Connor et~al\mbox{.}(2010)]%
        {o2010tweets}
\bibfield{author}{\bibinfo{person}{Brendan O'Connor}, \bibinfo{person}{Ramnath
  Balasubramanyan}, \bibinfo{person}{Bryan~R Routledge}, {and}
  \bibinfo{person}{Noah~A Smith}.} \bibinfo{year}{2010}\natexlab{}.
\newblock \showarticletitle{From tweets to polls: Linking text sentiment to
  public opinion time series}. In \bibinfo{booktitle}{\emph{Fourth
  international AAAI conference on weblogs and social media}}.
\newblock


\bibitem[Opgenhaffen and Scheerlinck(2014)]%
        {opgenhaffen2014social}
\bibfield{author}{\bibinfo{person}{Micha{\"e}l Opgenhaffen} {and}
  \bibinfo{person}{Harald Scheerlinck}.} \bibinfo{year}{2014}\natexlab{}.
\newblock \showarticletitle{Social media guidelines for journalists: An
  investigation into the sense and nonsense among Flemish journalists}.
\newblock \bibinfo{journal}{\emph{Journalism Practice}} \bibinfo{volume}{8},
  \bibinfo{number}{6} (\bibinfo{year}{2014}), \bibinfo{pages}{726--741}.
\newblock


\bibitem[Pedregosa et~al\mbox{.}(2011)]%
        {pedregosa2011scikit}
\bibfield{author}{\bibinfo{person}{Fabian Pedregosa}, \bibinfo{person}{Ga{\"e}l
  Varoquaux}, \bibinfo{person}{Alexandre Gramfort}, \bibinfo{person}{Vincent
  Michel}, \bibinfo{person}{Bertrand Thirion}, \bibinfo{person}{Olivier
  Grisel}, \bibinfo{person}{Mathieu Blondel}, \bibinfo{person}{Peter
  Prettenhofer}, \bibinfo{person}{Ron Weiss}, \bibinfo{person}{Vincent
  Dubourg}, {et~al\mbox{.}}} \bibinfo{year}{2011}\natexlab{}.
\newblock \showarticletitle{Scikit-learn: Machine learning in Python}.
\newblock \bibinfo{journal}{\emph{the Journal of machine Learning research}}
  \bibinfo{volume}{12} (\bibinfo{year}{2011}), \bibinfo{pages}{2825--2830}.
\newblock


\bibitem[Ranade et~al\mbox{.}(2013)]%
        {ranade2013stance}
\bibfield{author}{\bibinfo{person}{Sarvesh Ranade}, \bibinfo{person}{Rajeev
  Sangal}, {and} \bibinfo{person}{Radhika Mamidi}.}
  \bibinfo{year}{2013}\natexlab{}.
\newblock \showarticletitle{Stance classification in online debates by
  recognizing users’ intentions}. In \bibinfo{booktitle}{\emph{Proceedings of
  the SIGDIAL 2013 Conference}}. \bibinfo{pages}{61--69}.
\newblock


\bibitem[Reimers and Gurevych(2019)]%
        {reimers2019sentence}
\bibfield{author}{\bibinfo{person}{Nils Reimers} {and} \bibinfo{person}{Iryna
  Gurevych}.} \bibinfo{year}{2019}\natexlab{}.
\newblock \showarticletitle{Sentence-bert: Sentence embeddings using siamese
  bert-networks}.
\newblock \bibinfo{journal}{\emph{arXiv preprint arXiv:1908.10084}}
  (\bibinfo{year}{2019}).
\newblock


\bibitem[Samih and Darwish(2020)]%
        {samih2020few}
\bibfield{author}{\bibinfo{person}{Younes Samih} {and} \bibinfo{person}{Kareem
  Darwish}.} \bibinfo{year}{2020}\natexlab{}.
\newblock \showarticletitle{A few topical tweets are enough for effective
  user-level stance detection}.
\newblock \bibinfo{journal}{\emph{arXiv preprint arXiv:2004.03485}}
  (\bibinfo{year}{2020}).
\newblock


\bibitem[She and Chen(2014)]%
        {she2014tomoha}
\bibfield{author}{\bibinfo{person}{Jieying She} {and} \bibinfo{person}{Lei
  Chen}.} \bibinfo{year}{2014}\natexlab{}.
\newblock \showarticletitle{Tomoha: Topic model-based hashtag recommendation on
  twitter}. In \bibinfo{booktitle}{\emph{Proceedings of the 23rd international
  conference on World Wide Web}}. \bibinfo{pages}{371--372}.
\newblock


\bibitem[Shin et~al\mbox{.}(2022)]%
        {shin2022mask}
\bibfield{author}{\bibinfo{person}{Jieun Shin}, \bibinfo{person}{Aimei Yang},
  \bibinfo{person}{Wenlin Liu}, \bibinfo{person}{Hye Min~Kim},
  \bibinfo{person}{Alvin Zhou}, {and} \bibinfo{person}{Jingyi Sun}.}
  \bibinfo{year}{2022}\natexlab{}.
\newblock \showarticletitle{Mask-wearing as a partisan issue: Social identity
  and communication of party norms on social media among political elites}.
\newblock \bibinfo{journal}{\emph{Social Media+ Society}} \bibinfo{volume}{8},
  \bibinfo{number}{1} (\bibinfo{year}{2022}),
  \bibinfo{pages}{20563051221086233}.
\newblock


\bibitem[Small(2011)]%
        {small2011hashtag}
\bibfield{author}{\bibinfo{person}{Tamara~A Small}.}
  \bibinfo{year}{2011}\natexlab{}.
\newblock \showarticletitle{What the hashtag? A content analysis of Canadian
  politics on Twitter}.
\newblock \bibinfo{journal}{\emph{Information, communication \& society}}
  \bibinfo{volume}{14}, \bibinfo{number}{6} (\bibinfo{year}{2011}),
  \bibinfo{pages}{872--895}.
\newblock


\bibitem[Sobhani et~al\mbox{.}(2017)]%
        {sobhani2017dataset}
\bibfield{author}{\bibinfo{person}{Parinaz Sobhani}, \bibinfo{person}{Diana
  Inkpen}, {and} \bibinfo{person}{Xiaodan Zhu}.}
  \bibinfo{year}{2017}\natexlab{}.
\newblock \showarticletitle{A dataset for multi-target stance detection}. In
  \bibinfo{booktitle}{\emph{Proceedings of the 15th Conference of the European
  Chapter of the Association for Computational Linguistics: Volume 2, Short
  Papers}}. \bibinfo{pages}{551--557}.
\newblock


\bibitem[Taylor(2021)]%
        {derrick2021trump}
\bibfield{author}{\bibinfo{person}{Derrick~Bryson Taylor}.}
  \bibinfo{year}{2021}\natexlab{}.
\newblock \showarticletitle{George Floyd Protests: A Timeline}.
\newblock \bibinfo{journal}{\emph{The New York Times Company}}
  (\bibinfo{date}{Nov} \bibinfo{year}{2021}).
\newblock
\urldef\tempurl%
\url{https://www.nytimes.com/article/george-floyd-protests-timeline.html}
\showURL{%
\tempurl}


\bibitem[Van~Green and Tyson(2020)]%
        {van2020facts}
\bibfield{author}{\bibinfo{person}{Ted Van~Green} {and} \bibinfo{person}{Alec
  Tyson}.} \bibinfo{year}{2020}\natexlab{}.
\newblock \showarticletitle{facts about partisan reactions to COVID-19 in the
  US}.
\newblock \bibinfo{journal}{\emph{Pew Research Center}}  \bibinfo{volume}{2}
  (\bibinfo{year}{2020}).
\newblock


\bibitem[Vaswani et~al\mbox{.}(2017)]%
        {vaswani2017attention}
\bibfield{author}{\bibinfo{person}{Ashish Vaswani}, \bibinfo{person}{Noam
  Shazeer}, \bibinfo{person}{Niki Parmar}, \bibinfo{person}{Jakob Uszkoreit},
  \bibinfo{person}{Llion Jones}, \bibinfo{person}{Aidan~N Gomez},
  \bibinfo{person}{{\L}ukasz Kaiser}, {and} \bibinfo{person}{Illia
  Polosukhin}.} \bibinfo{year}{2017}\natexlab{}.
\newblock \showarticletitle{Attention is all you need}.
\newblock \bibinfo{journal}{\emph{Advances in neural information processing
  systems}}  \bibinfo{volume}{30} (\bibinfo{year}{2017}).
\newblock


\bibitem[Walker et~al\mbox{.}(2012)]%
        {walker2012stance}
\bibfield{author}{\bibinfo{person}{Marilyn Walker}, \bibinfo{person}{Pranav
  Anand}, \bibinfo{person}{Rob Abbott}, {and} \bibinfo{person}{Ricky Grant}.}
  \bibinfo{year}{2012}\natexlab{}.
\newblock \showarticletitle{Stance classification using dialogic properties of
  persuasion}. In \bibinfo{booktitle}{\emph{Proceedings of the 2012 conference
  of the North American chapter of the association for computational
  linguistics: Human language technologies}}. \bibinfo{pages}{592--596}.
\newblock


\bibitem[Wang et~al\mbox{.}(2012)]%
        {wang2012system}
\bibfield{author}{\bibinfo{person}{Hao Wang}, \bibinfo{person}{Do{\u{g}}an
  Can}, \bibinfo{person}{Abe Kazemzadeh}, \bibinfo{person}{Fran{\c{c}}ois Bar},
  {and} \bibinfo{person}{Shrikanth Narayanan}.}
  \bibinfo{year}{2012}\natexlab{}.
\newblock \showarticletitle{A system for real-time twitter sentiment analysis
  of 2012 us presidential election cycle}. In
  \bibinfo{booktitle}{\emph{Proceedings of the ACL 2012 system
  demonstrations}}. \bibinfo{pages}{115--120}.
\newblock


\bibitem[Wang et~al\mbox{.}(2019)]%
        {wang2019multi}
\bibfield{author}{\bibinfo{person}{Zhiguo Wang}, \bibinfo{person}{Patrick Ng},
  \bibinfo{person}{Xiaofei Ma}, \bibinfo{person}{Ramesh Nallapati}, {and}
  \bibinfo{person}{Bing Xiang}.} \bibinfo{year}{2019}\natexlab{}.
\newblock \showarticletitle{Multi-passage bert: A globally normalized bert
  model for open-domain question answering}.
\newblock \bibinfo{journal}{\emph{arXiv preprint arXiv:1908.08167}}
  (\bibinfo{year}{2019}).
\newblock


\bibitem[Weine et~al\mbox{.}(2020)]%
        {weine2020justice}
\bibfield{author}{\bibinfo{person}{Stevan Weine}, \bibinfo{person}{Brandon~A
  Kohrt}, \bibinfo{person}{Pamela~Y Collins}, \bibinfo{person}{Janice Cooper},
  \bibinfo{person}{Roberto Lewis-Fernandez}, \bibinfo{person}{Samuel Okpaku},
  {and} \bibinfo{person}{Milton~L Wainberg}.} \bibinfo{year}{2020}\natexlab{}.
\newblock \showarticletitle{Justice for George Floyd and a reckoning for global
  mental health}.
\newblock \bibinfo{journal}{\emph{Global Mental Health}}  \bibinfo{volume}{7}
  (\bibinfo{year}{2020}).
\newblock


\bibitem[Yang(2016)]%
        {yang2016narrative}
\bibfield{author}{\bibinfo{person}{Guobin Yang}.}
  \bibinfo{year}{2016}\natexlab{}.
\newblock \showarticletitle{Narrative agency in hashtag activism: The case of\#
  BlackLivesMatter}.
\newblock \bibinfo{journal}{\emph{Media and communication}}
  \bibinfo{volume}{4}, \bibinfo{number}{4} (\bibinfo{year}{2016}),
  \bibinfo{pages}{13}.
\newblock


\bibitem[Young et~al\mbox{.}(2022)]%
        {young2022politics}
\bibfield{author}{\bibinfo{person}{Dannagal~G Young}, \bibinfo{person}{Huma
  Rasheed}, \bibinfo{person}{Amy Bleakley}, {and} \bibinfo{person}{Jessica~B
  Langbaum}.} \bibinfo{year}{2022}\natexlab{}.
\newblock \showarticletitle{The politics of mask-wearing: Political
  preferences, reactance, and conflict aversion during COVID}.
\newblock \bibinfo{journal}{\emph{Social Science \& Medicine}}
  (\bibinfo{year}{2022}), \bibinfo{pages}{114836}.
\newblock


\bibitem[Zhang et~al\mbox{.}(2020)]%
        {zhang2020semantics}
\bibfield{author}{\bibinfo{person}{Zhuosheng Zhang}, \bibinfo{person}{Yuwei
  Wu}, \bibinfo{person}{Hai Zhao}, \bibinfo{person}{Zuchao Li},
  \bibinfo{person}{Shuailiang Zhang}, \bibinfo{person}{Xi Zhou}, {and}
  \bibinfo{person}{Xiang Zhou}.} \bibinfo{year}{2020}\natexlab{}.
\newblock \showarticletitle{Semantics-aware BERT for language understanding}.
  In \bibinfo{booktitle}{\emph{Proceedings of the AAAI Conference on Artificial
  Intelligence}}, Vol.~\bibinfo{volume}{34}. \bibinfo{pages}{9628--9635}.
\newblock


\end{thebibliography}
%%% -*-BibTeX-*-
%%% Do NOT edit. File created by BibTeX with style
%%% ACM-Reference-Format-Journals [18-Jan-2012].

\appendix

\section{Data Collection Keywords}
\label{app:data_collection_keywords}

\begin{itemize}[leftmargin=*]
    \item \textbf{Election Data Keywords}: election, presidential election, election candidate, \#2020election, \#vote2020, \#election2020, \#elections2020, \#debates, \#debate, \#debatenight, presidentialdebate, joe, biden,  joebiden, joe biden, bernie, sanders, berniesanders, bernie sanders, warren, elizabeth, elizabethwarren, \#votebluenomatterwho, \#votebluetosaveamerica, \#votethemallout2020, \#votebluenomatterwho2020, \#nevertrump, \#democrats, donald, trump,  donaldtrump, donald trump, \#trump2020, \#maga, \#tcot, \#p2, \#republicanparty, \#gop, kamala, harris, kamalaharris, kamala harris
    \item \textbf{COVID-19 Data keywords}: coronavirus, covid, chinese virus, wuhan, ncov, sars-cov-2, koronavirus, corona, cdc, N95, kungflu, epidemic, outbreak, sinophobia, china, pandemic, covid
    \item \textbf{Racial Data Equality keywords}: george floyd, race, \#blacklivesmatter, racial, racist, racism, nigga, nigger, black america, white american, black people, white people, black lives matter, black man, white man, black men, white men, black woman, white woman, black women, white women, \#georgefloyd, \#blm, blackman, \#whiteman, \#blackmen, \#whitemen, \#blackwoman, \#whitewoman, \#blackwomen, \#whitewomen, \#blackamerican, \#whiteamerican, \#blackpeople, \#whitepeople
\end{itemize}

\section{Hashtag Based Stance Detector Algorithm}
\label{app: hashtag_stance_detector_algo}

The algorithm 1 describes running hashtag-based stance detector on tweet $t$ with Support hashtags $S$ and Against hashtags $A$.

\begin{algorithm}[b]
\caption{Hashtag Based Stance Detector}\label{alg:one}
\KwIn{$t, S, A$}
\KwOut{$stance$}

    \SetKwFunction{FMain}{detectStance}
    \SetKwProg{Fn}{Function}{:}{}
    \Fn{\FMain{t, S, A}}{
        $ score \longleftarrow 0 $\;
        \ForEach{hashtag in S}
        {\If{hashtag exist in t}
            {$score \longleftarrow score + 1$}
        }    
        \ForEach{hashtag in A}
        {{\If{hashtag exist in t}
            {$score \longleftarrow score - 1$}
        }
        \uIf{score > 0}{
            \Return Support \;
        }
        \uElseIf{score = 0}{
            \Return Non-Opinionated \;
        }
        \Else{
            \Return Against \;
        }
        }
}
\textbf{End Function}
\end{algorithm}

\end{document}